\documentclass[10pt,a4paper,twocolumn,english]{IEEEtran}
\usepackage[T1]{fontenc}
\usepackage{verbatim}
\usepackage{float}
\usepackage{mathrsfs}
\usepackage{amsmath}
\usepackage{amsthm}
\usepackage{amssymb}
\usepackage{graphicx}
\usepackage{setspace}

\makeatletter

\pdfpageheight\paperheight
\pdfpagewidth\paperwidth

\floatstyle{ruled}
\newfloat{algorithm}{tbp}{loa}
\providecommand{\algorithmname}{Algorithm}
\floatname{algorithm}{\protect\algorithmname}

\theoremstyle{plain}
\newtheorem{thm}{\protect\theoremname}
\theoremstyle{plain}
\newtheorem{lem}[thm]{\protect\lemmaname}
\theoremstyle{plain}
\newtheorem{prop}[thm]{\protect\propositionname}

\usepackage{subfigure}
\usepackage{epstopdf}
\usepackage{cite}
\usepackage{citesort}
\usepackage{balance}

\makeatother

\usepackage{babel}
\providecommand{\lemmaname}{Lemma}
\providecommand{\propositionname}{Proposition}
\providecommand{\theoremname}{Theorem}

\begin{document}

\title{Performance Impact of Idle Mode Capability on Dense Small Cell Networks
}
\begin{singlespace}

\author{\noindent {\normalsize{}Ming Ding, }\textit{\normalsize{}Senior Member,
IEEE}{\normalsize{}, David L$\acute{\textrm{o}}$pez P$\acute{\textrm{e}}$rez,
}\textit{\normalsize{}Senior Member, IEEE}{\normalsize{}, }\\
{\normalsize{}Guoqiang Mao, }\textit{\normalsize{}Senior Member, IEEE}{\normalsize{},
Zihuai Lin, }\textit{\normalsize{}Senior Member, IEEE}
}
\end{singlespace}
\maketitle
\begin{abstract}
Very recent studies showed that in a fully loaded dense small cell
network (SCN), the coverage probability performance will continuously
\emph{decrease} with the network densification. Such new results were
captured in IEEE ComSoc Technology News with an alarming title of
``Will Densification Be the Death of 5G?''. In this paper, we revisit
this issue from more practical views of realistic network deployment,
such as a finite number of active base stations (BSs) and user equipments
(UEs), a decreasing BS transmission power with the network densification,
and so on. Particularly, in dense SCNs, due to an oversupply of BSs
with respect to UEs, a large number of BSs can be put into idle modes
without signal transmission, if there is no active UE within their
coverage areas. Setting those BSs into idle modes mitigates unnecessary
inter-cell interference and reduces energy consumption. In this paper,
we investigate the performance impact of such BS idle mode capability
(IMC) on dense SCNs. Different from existing work, we consider a realistic
path loss model incorporating both line-of-sight (LoS) and non-line-of-sight
(NLoS) transmissions. Moreover, we obtain analytical results for the
coverage probability, the area spectral efficiency (ASE) and the energy
efficiency (EE) performance for SCNs with the BS IMC and show that
the performance impact of the IMC on dense SCNs is significant. As
the BS density surpasses the UE density in dense SCNs, the coverage
probability will continuously \emph{increase toward one}, addressing
previous concerns on ``the death of 5G''. Finally, the performance
improvement in terms of the EE performance is also investigated for
dense SCNs using practical energy models developed in the Green-Touch
project.%
\footnote{To appear in IEEE TVT. 1536-1276 © 2015 IEEE. Personal use is permitted, but republication/redistribution requires IEEE permission. Please find the final version in IEEE from the link: http://ieeexplore.ieee.org/document/xxxxxxx/. Digital Object Identifier: 10.1109/TVT.2017.xxxxxxx}
\end{abstract}


\section{Introduction\label{sec:Introduction}}

Dense small cell networks (SCNs), comprised of remote radio heads,
metrocells, picocells, femtocells, relay nodes, etc., have attracted
significant attention as one of the most promising approaches to rapidly
increase network capacity and meet the ever-increasing data traffic
demands~\cite{Report_CISCO}. Indeed, the orthogonal deployment of
dense SCNs within the existing macrocell networks~\cite{TR36.872},
i.e., small cells and macrocells operating on different frequency
spectrum (Small Cell Scenario \#2a~\cite{TR36.872}), has been selected
as the workhorse for capacity enhancement in the 4th-generation (4G)
and the 5th-generation (5G) networks, developed by the 3rd Generation
Partnership Project (3GPP)~\cite{Tutor_smallcell}.%
{} In this paper, we focus on the analysis of these dense SCNs with
an orthogonal deployment in the existing macrocell networks.

\begin{figure}
\noindent \begin{centering}
\includegraphics[width=8cm]{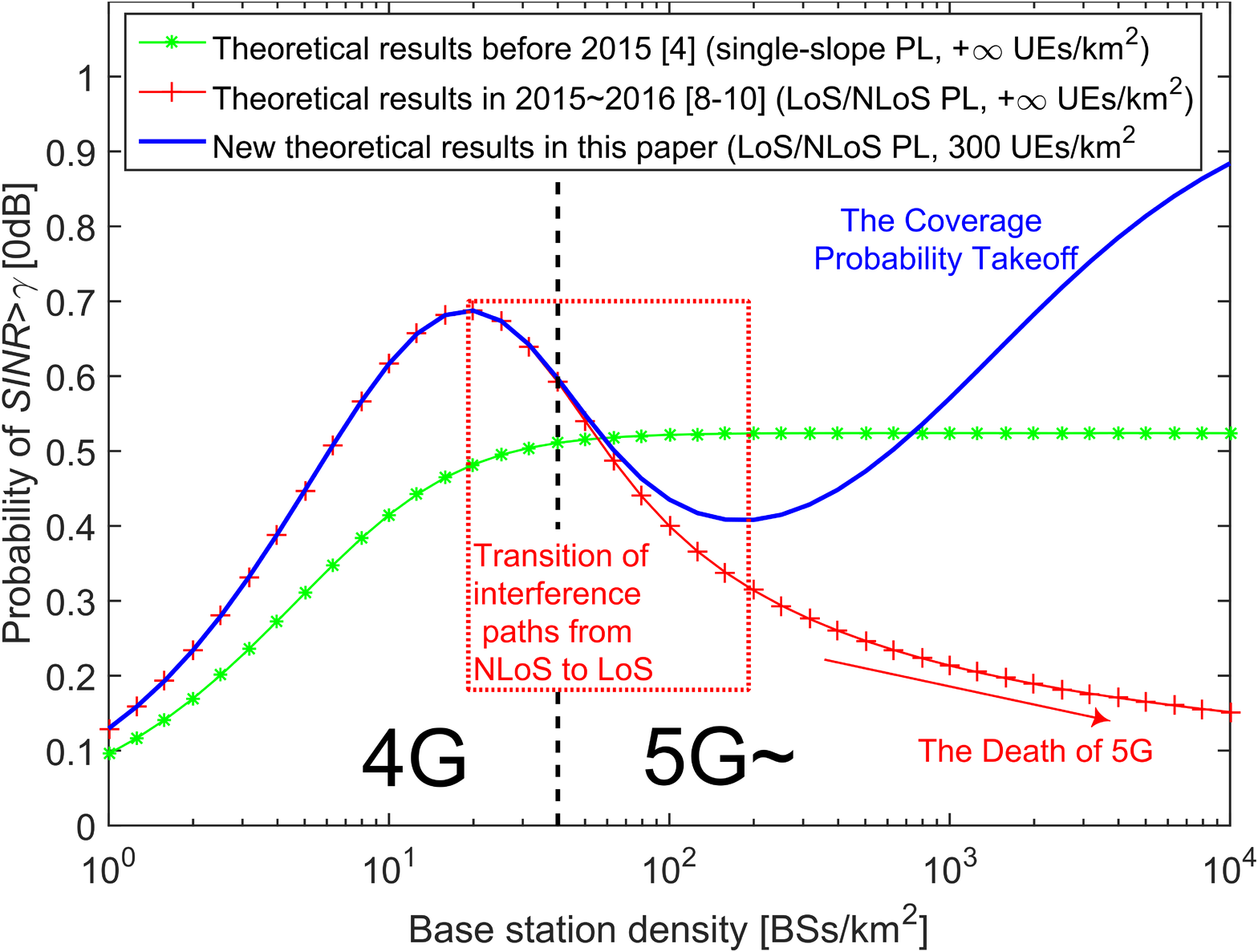}\renewcommand{\figurename}{Fig.}\caption{\label{fig:comp_p_cov_4Gto5G}Theoretical performance comparison of
the coverage probability when the SINR threshold $\gamma=0$\,dB.
Note that all the results are obtained using practical 3GPP channel
models~\cite{TR36.828,SCM_pathloss_model}, which will be introduced
in details later. Moreover, the BS density regions for the 4G and
5G networks have been illustrated in the figure, considering that
the typical BS density of the 4G SCNs is in the order of tens of $\textrm{BSs/km}^{2}$~\cite{TR36.872,Tutor_smallcell}. }
\par\end{centering}
\vspace{-0.5cm}
\end{figure}

In the seminal work of Andrews, Baccelli, and Ganti~\cite{Jeff2011},
a conclusion was reached: the density of base stations (BSs) would
not affect the coverage probability performance in interference-limited\footnote{In a interference-limited network, the power of each BS is set to
a value much larger than the noise power. } and fully-loaded\footnote{In a fully-loaded network, all BSs are active. Such assumption implies
that the user density is infinity or much larger than the BS density.
According to the results in~\cite{dynOnOff_Huang2012}, the user
density should be at least 10 times higher than the BS density to
make sure that almost all BSs are active. } wireless networks, where the coverage probability is defined as the
probability that the signal-to-interference-plus-noise ratio (SINR)
of a typical user equipment (UE) is above a SINR threshold $\gamma$.
Consequently, the area spectral efficiency (ASE) performance in $\textrm{bps/Hz/km}^{2}$
would scale linearly with the network densification~\cite{Jeff2011},
which forecasts a bright future for dense SCNs in 4G and 5G.%
{} The intuition of such conclusion is that the increase in the interference
power caused by a denser network would be exactly compensated by the
increase in the signal power due to the reduced distance between transmitters
and receivers. This coverage probability behavior predicted in~\cite{Jeff2011}
is shown in Fig.~\ref{fig:comp_p_cov_4Gto5G}.
{} However, it is important to note that such conclusion was obtained
with considerable simplifications on the network condition and propagation
environment. For example, all BSs were assumed to be active and a
single-slop path loss model was used. It would be interesting to investigate
whether the conclusion still holds in real-world environment featuring
more complicated BS behaviors and radio propagation environment.

To this end, a few noteworthy studies have been carried out recently
to revisit the network performance analysis of dense SCNs using more
practical propagation models. In~\cite{related_work_Jeff}, the authors
considered a multi-slope piece-wise path loss function, while in~\cite{Related_work_Health},
the authors modeled line-of-sight (LoS) and non-line-of-sight (NLoS)
transmissions%
{} as probabilistic events for a millimeter wave communication scenario.
In our recent work~\cite{our_work_TWC2016}, we further considered
both piece-wise path loss functions and probabilistic LoS and NLoS
transmissions. The above new studies demonstrated that when the BS
density is larger than a threshold $\lambda^{*}$, the coverage probability
will continuously \emph{decrease} as the SCN becomes denser. The intuition
behind such result is that the interference power increases faster
than the signal power in dense SCNs due to the transition of a large
number of interference paths from NLoS (usually with a large path
loss exponent) to LoS (usually with a small path loss exponent). Such
new results were later captured in IEEE ComSoc Technology News with
an alarming title of ``Will Densification Be the Death of 5G?''~\cite{death5G2015News}.%
{} Fig.~\ref{fig:comp_p_cov_4Gto5G} shows the coverage probability
result in~\cite{related_work_Jeff,Related_work_Health,our_work_TWC2016},
where $\lambda^{*}$ is around 20\,$\textrm{BSs/km}^{2}$.%
{} The key message is that, when deploying dense SCNs, an increased
BS density may lead to worse network performance, and hence the future
of 5G is shrouded in darkness.%

In this paper, we will take another look at ``the death of 5G''
from more practical views of realistic network deployment, such as
a finite number of active BSs and UEs, a decreasing BS transmission
power with the network densification, and so on. Particularly, since
the UE density is finite in practical networks, a large number of
BSs in dense SCNs could switch off their transmission modules and
thus enter idle modes, if there is no active UE within their coverage
areas. Setting those BSs to idle modes can mitigate unnecessary inter-cell
interference and reduce energy consumption~\cite{dynOnOff_Huang2012,Luo2014_dynOnOff,Li2014_energyEff,Zhang2016HetNetEE}.
In other words, by dynamically muting idle BSs, the interference suffered
by UEs from always-on control channels, e.g., synchronization and
broadcast channels, and data channels can be reduced, thus improving
UEs' coverage probability. This idle mode feature at BSs is referred
to as the idle mode capability (IMC) hereafter. Furthermore, the energy
efficiency (EE) of SCNs with the IMC can be significantly enhanced
because (i) BSs without any active UE can be temporarily put into
idle modes with low energy consumption, and (ii) every active BS usually
benefits from high-SINR and thus energy-efficient links with its associated
UEs due to the BS diversity gain~\cite{dynOnOff_Huang2012}, i.e.,
each UE selects the serving BS with the highest SINR from a surplus
of BSs in dense SCNs. It is very important to note that a BS in idle
mode may still consume a non-negligible amount of energy, thus impacting
the EE of SCNs. In this paper, we use a practical power model developed
in the Green-Touch project~\cite{DessetPowerModelling} to evaluate
the EE performance in realistic scenarios. Such power model will be
formally introduced later.

In this paper, we investigate for the first time the performance impact
of the IMC on dense SCNs considering LoS and NLoS transmissions. As
an example to demonstrate such impact, our results with a UE density
of $300\thinspace\textrm{UEs/km}^{2}$ (a typical UE density in 5G~\cite{Tutor_smallcell})
are compared with the existing results in Fig.~\ref{fig:comp_p_cov_4Gto5G}.
The performance impact of the IMC on the coverage probability is shown
to be significant. As the BS density surpasses the UE density in future
dense and ultra-dense SCNs~\cite{Tutor_smallcell}, thus creating
a surplus of BSs, the coverage probability will continuously\emph{
increase toward one}, addressing the critical issue of coverage probability
decrease that may cause ``the death of 5G'' shown in Fig.~\ref{fig:comp_p_cov_4Gto5G}.
Such performance behavior of the coverage probability increasing toward
one in dense SCNs, is referred to as\emph{ the Coverage Probability
Takeoff} hereafter. The intuition behind \emph{the Coverage Probability
Takeoff} is that beyond a certain BS density threshold, the interference
power will be less than that of the case with all BSs being active
thanks to the BS IMC, plus the signal power will continuously rise
due to the BS diversity gain, thus leading to a better SINR performance
as the network evolves into a dense one.

Compared with existing work, the main contributions of this paper
are\footnote{Note that preliminary results of this work has been presented in a
conference paper~\cite{Ding2016IMC_GC}.}:
\begin{itemize}
\item Analytical results are obtained for the coverage probability and the
ASE performance of SCNs with the BS IMC using a general path loss
model incorporating both LoS and NLoS transmissions. Note that existing
work on the IMC only treated a single-slope path loss model, where
a UE is always associated with its nearest BS~\cite{dynOnOff_Huang2012,Li2014_energyEff},
while our work considers more practical path loss models with probabilistic
LoS and NLoS transmissions, where UEs may connect to a farther BS
with a LoS path.%
\item A lower bound, an upper bound and an approximate expression of the
active BS density are derived for SCNs with the IMC, considering practical
path loss models with probabilistic LoS and NLoS transmissions.
\item The performance improvement in terms of the EE is also investigated
for dense SCNs using practical energy models developed in the Green-Touch
project~\cite{DessetPowerModelling} and practical 3GPP propagation
models with \emph{Rician fading}, \emph{correlated shadow fading},
etc.%
\end{itemize}

The rest of this paper is structured as follows. Section~\ref{sec:Related-Work}
provides a brief review of related work. Section~\ref{sec:System-Model}
describes the system model featuring the BS IMC. Section~\ref{sec:Main-Results}
presents our theoretical results on the coverage probability, the
ASE, the EE and the active BS density, with their applications in
two 3GPP special cases. The numerical results are discussed in Section~\ref{sec:Simulation-and-Discussion},
with remarks shedding new light on the issue of ``the death of 5G''.%
{} The conclusions are drawn in Section~\ref{sec:Conclusion}.

\section{Related Work\label{sec:Related-Work}}

In stochastic geometry, BS positions are typically modeled as a Homogeneous
Poisson Point Process (HPPP) on the plane, and closed-form expressions
of coverage probability can be found for some scenarios in single-tier
cellular networks~\cite{Jeff2011} and multi-tier cellular networks~\cite{Dhillon2012hetNetSG}.
The major conclusion in~\cite{Jeff2011,Dhillon2012hetNetSG} is that
neither the number of cells nor the number of cell tiers changes the
coverage probability in interference-limited fully-loaded wireless
networks.

Recently, a few noteworthy studies have been carried out to further
investigate the network performance analysis for dense and ultra-dense
SCNs under more practical propagation models. As discussed in Section~\ref{sec:Introduction},
the authors of~\cite{related_work_Jeff,Related_work_Health,our_work_TWC2016}
found that the coverage probability performance will start to decrease
when the BS density is sufficiently large%
. %
The intuition behind this result is that as the BS density becomes
larger than a threshold, the interference power increases faster than
the signal power due to the transition of a large number of interference
paths from NLoS to LoS.%
{}

However, all of the above work did not consider an important factor:
as the BS density increases, a large number of BSs can be put into
idle mode without signal transmission, if there is no active UE within
their coverage areas. This is a new network behavior arising from
the surplus of BSs with respect to UEs, i.e., it may happen that a
significant number of BSs may not have any active UE in their coverage
areas during certain time periods. Therefore, such BSs could mute
their transmission to mitigate unnecessary inter-cell interference
and reduce energy consumption~\cite{dynOnOff_Huang2012,Luo2014_dynOnOff,Li2014_energyEff,Renzo2016intensityMatch,Zhang2016HetNetEE}.

Up to now, the limited existing work that did consider the IMC, only
treated a simplistic single-slope path loss model for homogeneous
SCNs~\cite{dynOnOff_Huang2012,Luo2014_dynOnOff,Li2014_energyEff,Renzo2016intensityMatch}
or for the co-channel deployment of heterogeneous networks~\cite{Zhang2016HetNetEE}.
Such path loss assumption is not practical for realistic SCNs and
may yield misleading conclusions reading the network performance,
as addressed in~\cite{related_work_Jeff,Related_work_Health,our_work_TWC2016}.%

Motivated by the above observations, in this paper, we investigate
for the first time the performance impact of the IMC on dense SCNs
considering probabilistic LoS and NLoS transmissions. Note that compared
with our previous work that also considered probabilistic LoS and
NLoS transmissions~\cite{our_work_TWC2016}, this paper present new
contributions as follows,
\begin{itemize}
\item Our previous work~\cite{our_work_TWC2016} corroborates \textquotedblleft the
death of 5G\textquotedblright ~\cite{death5G2015News} by considering
probabilistic LoS and NLoS transmissions and an infinite number of
UEs in the network. However, in this work, we present new theoretical
results that can mitigate \textquotedblleft the death of 5G\textquotedblright{}
by considering a finite number of UEs exploited by the BS IMC.
\item The new theoretical work in this paper compared with~\cite{our_work_TWC2016}
is that a lower bound, an upper bound and an approximate expression
of the active BS density are derived for SCNs with the IMC, considering
practical path loss models with probabilistic LoS and NLoS transmissions.
\item Moreover, compared with~\cite{our_work_TWC2016}, the performance
improvement in terms of the EE is also investigated in this paper.
\end{itemize}

\section{System Model\label{sec:System-Model}}

We consider a downlink (DL) cellular network with BSs deployed on
a plane according to a homogeneous Poisson point process (HPPP) $\Phi$
with a density of $\lambda$ $\textrm{BSs/km}^{2}$. Active UEs are
Poisson distributed in the considered DL network with a density of
$\rho$ $\textrm{UEs/km}^{2}$. Here, we only consider active UEs
in the network because non-active UEs do not trigger data transmission,
and thus they are ignored in our analysis. Note that the total UE
number in cellular networks should be much higher than the number
of the active UEs, but at a certain time slot and on a certain frequency
band, the active UEs with data traffic demands are not too many. As
discussed in Section~\ref{sec:Introduction}, a typical density of
the active UEs in 5G should be around $300\thinspace\textrm{UEs/km}^{2}$~\cite{Tutor_smallcell}.

In our previous work~\cite{our_work_TWC2016,our_GC_paper_2015_HPPP}
and other related work~\cite{related_work_Jeff,Related_work_Health},
$\rho$ was assumed to be infinite or considerably larger than $\lambda$
so that each BS has at least one associated UE in its coverage. In
this work, we impose no such constraint on $\rho$, and hence a BS
with the IMC will enter an idle mode if there is no UE connected to
it, which reduces interference to neighboring UEs as well as energy
consumption of the network. Since UEs are randomly and uniformly distributed
in the network, we assume that the active BSs also follow an HPPP
distribution $\tilde{\Phi}$~\cite{dynOnOff_Huang2012,Luo2014_dynOnOff,Li2014_energyEff,Renzo2016intensityMatch},
the density of which is denoted by $\tilde{\lambda}$ $\textrm{BSs/km}^{2}$.
Note that $\tilde{\lambda}\leq\lambda$, and $\tilde{\lambda}\leq\rho$
since one UE is served by at most one BS. Obviously, a larger $\rho$
requires more active BSs with a larger $\tilde{\lambda}$ to serve
the active UEs.

It is very important to note that, up to now, there is no theoretical
proof showing that the active BSs should follow an HPPP since the
activation of each BS depends on the UE distribution in its vicinity.
Having said that, the HPPP assumption has been widely used in the
literature, such as~\cite{dynOnOff_Huang2012,Luo2014_dynOnOff,Li2014_energyEff,Renzo2016intensityMatch}.
Indeed, later we will present simulation results backing up our theoretical
findings based on the HPPP assumption, where the computational engines
for the computer simulations and theoretical analyses follow different
principles. More specifically, in our simulations, no assumption was
made on the distribution of the active BSs. They are generated according
to the UEs\textquoteright{} selection. In contrast, the HPPP assumption
was only used to obtain the analytical results. This methodology has
also been used in~\cite{dynOnOff_Huang2012,Luo2014_dynOnOff,Li2014_energyEff,Renzo2016intensityMatch}.
The intuition of this conclusion is that since no clustering behavior
of UEs and no correlation among UEs\textquoteright{} channels have
been considered in the analysis, the activation and deactivation of
each BS is uniformly and randomly distributed across the entire network,
which leads to the HPPP assumption.

Following~\cite{our_GC_paper_2015_HPPP,our_work_TWC2016}, we adopt
a very general path loss model, in which the path loss $\zeta\left(r\right)$
as a function of $r$ is segmented into $N$ pieces written as%
\begin{equation}
\zeta\left(r\right)=\begin{cases}
\zeta_{1}\left(r\right), & \textrm{when }0\leq r\leq d_{1}\\
\zeta_{2}\left(r\right), & \textrm{when }d_{1}<r\leq d_{2}\\
\vdots & \vdots\\
\zeta_{N}\left(r\right), & \textrm{when }r>d_{N-1}
\end{cases},\label{eq:prop_PL_model}
\end{equation}
where each piece $\zeta_{n}\left(r\right),n\in\left\{ 1,2,\ldots,N\right\} $
is modeled as
\begin{equation}
\zeta_{n}\left(r\right)\hspace{-0.1cm}=\hspace{-0.1cm}\begin{cases}
\hspace{-0.2cm}\begin{array}{l}
\zeta_{n}^{\textrm{L}}\left(r\right)=A_{n}^{{\rm {L}}}r^{-\alpha_{n}^{{\rm {L}}}},\\
\zeta_{n}^{\textrm{NL}}\left(r\right)=A_{n}^{{\rm {NL}}}r^{-\alpha_{n}^{{\rm {NL}}}},
\end{array} & \hspace{-0.2cm}\hspace{-0.3cm}\begin{array}{l}
\textrm{LoS Prob.:}~\textrm{Pr}_{n}^{\textrm{L}}\left(r\right)\\
\textrm{NLoS Prob.:}~1-\textrm{Pr}_{n}^{\textrm{L}}\left(r\right)
\end{array}\hspace{-0.1cm},\end{cases}\label{eq:PL_BS2UE}
\end{equation}
where
\begin{itemize}
\item $\zeta_{n}^{\textrm{L}}\left(r\right)$ and $\zeta_{n}^{\textrm{NL}}\left(r\right),n\in\left\{ 1,2,\ldots,N\right\} $
are the $n$-th piece path loss functions for the LoS transmission
and the NLoS transmission, respectively,
\item $A_{n}^{{\rm {L}}}$ and $A_{n}^{{\rm {NL}}}$ are the path losses
at a reference distance $r=1$ for the LoS and the NLoS cases, respectively,
\item $\alpha_{n}^{{\rm {L}}}$ and $\alpha_{n}^{{\rm {NL}}}$ are the path
loss exponents for the LoS and the NLoS cases, respectively.
\end{itemize}
\noindent In practice, $A_{n}^{{\rm {L}}}$, $A_{n}^{{\rm {NL}}}$,
$\alpha_{n}^{{\rm {L}}}$ and $\alpha_{n}^{{\rm {NL}}}$ are constants
obtainable from field tests~\cite{TR36.828,SCM_pathloss_model}.

Moreover, $\textrm{Pr}_{n}^{\textrm{L}}\left(r\right)$ is the $n$-th
piece LoS probability function that a transmitter and a receiver separated
by a distance $r$ has a LoS path, which is assumed to be \emph{a
monotonically decreasing function} with regard to $r$. Such assumption
has been confirmed by existing measurement studies~\cite{TR36.828,SCM_pathloss_model}.
For convenience, $\left\{ \zeta_{n}^{\textrm{L}}\left(r\right)\right\} $
and $\left\{ \zeta_{n}^{\textrm{NL}}\left(r\right)\right\} $ are
further stacked into piece-wise functions written as
\begin{equation}
\zeta^{Path}\left(r\right)=\begin{cases}
\zeta_{1}^{Path}\left(r\right), & \textrm{when }0\leq r\leq d_{1}\\
\zeta_{2}^{Path}\left(r\right),\hspace{-0.3cm} & \textrm{when }d_{1}<r\leq d_{2}\\
\vdots & \vdots\\
\zeta_{N}^{Path}\left(r\right), & \textrm{when }r>d_{N-1}
\end{cases},\label{eq:general_PL_func}
\end{equation}
where the string variable $Path$ takes the value of ``L'' and ``NL''
for the LoS and the NLoS cases, respectively.

Besides, $\left\{ \textrm{Pr}_{n}^{\textrm{L}}\left(r\right)\right\} $
is stacked into a piece-wise function as
\begin{equation}
\textrm{Pr}^{\textrm{L}}\left(r\right)=\begin{cases}
\textrm{Pr}_{1}^{\textrm{L}}\left(r\right), & \textrm{when }0\leq r\leq d_{1}\\
\textrm{Pr}_{2}^{\textrm{L}}\left(r\right),\hspace{-0.3cm} & \textrm{when }d_{1}<r\leq d_{2}\\
\vdots & \vdots\\
\textrm{Pr}_{N}^{\textrm{L}}\left(r\right), & \textrm{when }r>d_{N-1}
\end{cases}.\label{eq:general_LoS_Pr}
\end{equation}

Note that the generality and the practicality of the adopted path
loss model (\ref{eq:prop_PL_model}) have been well established in~\cite{our_work_TWC2016}.
In more detail, this model is consistent with the ones adopted in
the 3GPP~\cite{TR36.828},~\cite{SCM_pathloss_model}, and includes
those models considered in~\cite{related_work_Jeff} and~\cite{Related_work_Health}
as its special cases.%

In this paper, we assume a practical user association strategy (UAS),
in which each UE is connected to the BS with the smallest path loss
(i.e., with the largest $\zeta\left(r\right)$) to the UE~\cite{Related_work_Health,our_work_TWC2016}.
Note that in our previous work~\cite{our_GC_paper_2015_HPPP} and
some other existing work, e.g.,~\cite{Jeff2011,related_work_Jeff},
it was assumed that each UE should be associated with its closest
BS. Such assumption is not appropriate for the considered path loss
model in (\ref{eq:prop_PL_model}), because in practice a UE should
connect to a BS offering the largest received signal strength. Such
BS does not necessarily have to be the nearest one to the UE, and
it could be a farther one with a strong LoS path.

Moreover, we assume that each BS/UE is equipped with an isotropic
antenna, and that the multi-path fading between a BS and a UE is modeled
as independently identical distributed (i.i.d.) Rayleigh fading~\cite{related_work_Jeff,Related_work_Health,our_work_TWC2016}.
Note that a practical 3GPP model with \emph{distance-dependent Rician
fading}~\cite{SCM_pathloss_model} and \emph{correlated shadow fading}~\cite{TR36.828}
will also be considered and simulated in Section~\ref{sec:Simulation-and-Discussion}
to show their minor impact on our conclusions. More specifically,
\begin{itemize}
\item We adopt a practical Rician fading defined in the 3GPP~\cite{SCM_pathloss_model},
where the $K$ factor in dB scale (the ratio between the power in
the direct path and the power in the other scattered paths) is modeled
as $K[\textrm{dB}]=13-0.03r$, where $r$ is the distance in meter.
\item We consider a practical correlated shadow fading defined in 3GPP~\cite{TR36.828},
where the shadow fading in dB is modeled as zero-mean Gaussian random
variables, e.g., with a standard deviation of 10$\,$dB. The correlation
coefficient between the shadow fading values associated with two different
BSs is denoted by $\tau$, e.g., $\tau=0.5$ in~\cite{TR36.828}.
\end{itemize}

\section{Main Results\label{sec:Main-Results}}

In this section, we study the performance of SCNs in terms of the
coverage probability, the ASE and the EE by considering the performance
of a typical UE located at the origin $o$.

\subsection{The Coverage Probability\label{subsec:The-Coverage-Probability}}

First, we investigate the coverage probability that the typical UE's
SINR is above a designated threshold $\gamma$:
\begin{equation}
p^{\textrm{cov}}\left(\lambda,\gamma\right)=\textrm{Pr}\left[\mathrm{SINR}>\gamma\right],\label{eq:Coverage_Prob_def}
\end{equation}
where the SINR is computed by
\begin{equation}
\mathrm{SINR}=\frac{P\zeta\left(r\right)h}{I_{\textrm{agg}}+P_{{\rm {N}}}}.\label{eq:SINR}
\end{equation}
Here, $h$ is the channel gain, which is modeled as an exponentially
distributed random variable (RV) with a mean of one (due to our consideration
of Rayleigh fading mentioned above), $P$ and $P_{{\rm {N}}}$ are
the BS transmission power and the additive white Gaussian noise (AWGN)
power at each UE, respectively, and $I_{\textrm{agg}}$ is the cumulative
interference given by
\begin{equation}
I_{\textrm{agg}}=\sum_{i:\,b_{i}\in\tilde{\Phi}\setminus b_{o}}P\beta_{i}g_{i},\label{eq:cumulative_interference}
\end{equation}
where $b_{o}$ is the BS serving the typical UE, and $b_{i}$, $\beta_{i}$
and $g_{i}$ are the $i$-th interfering BS, the path loss from $b_{i}$
to the typical UE and the multi-path fading channel gain associated
with $b_{i}$, respectively. Note that when all BSs are assumed to
be active, the set of all BSs $\Phi$ should be used in the expression
of $I_{\textrm{agg}}$~\cite{related_work_Jeff,Related_work_Health,our_work_TWC2016}.
Here, in (\ref{eq:cumulative_interference}), only the active BSs
in $\tilde{\Phi}\setminus b_{o}$ inject effective interference into
the network, where $\tilde{\Phi}$ denotes the set of the active BSs.
In other words, the BSs in idle modes are not taken into account in
the analysis of $I_{\textrm{agg}}$.

Based on the path loss model in (\ref{eq:prop_PL_model}) and the
adopted UAS, we present our result of $p^{\textrm{cov}}\left(\lambda,\gamma\right)$
in Theorem~\ref{thm:p_cov_UAS1}.%
{\small{}}
\begin{algorithm*}
\begin{thm}
\begin{doublespace}
\label{thm:p_cov_UAS1}Considering the path loss model in (\ref{eq:prop_PL_model})
and the presented UAS, the probability of coverage $p^{{\rm {cov}}}\left(\lambda,\gamma\right)$
can be derived as
\begin{equation}
p^{{\rm {cov}}}\left(\lambda,\gamma\right)=\sum_{n=1}^{N}\left(T_{n}^{{\rm {L}}}+T_{n}^{{\rm {NL}}}\right),\label{eq:Theorem_1_p_cov}
\end{equation}
where $T_{n}^{{\rm {L}}}=\int_{d_{n-1}}^{d_{n}}{\rm {Pr}}\left[\frac{P\zeta_{n}^{{\rm {L}}}\left(r\right)h}{I_{{\rm {agg}}}+P_{{\rm {N}}}}>\gamma\right]f_{R,n}^{{\rm {L}}}\left(r\right)dr$,
$T_{n}^{{\rm {NL}}}=\int_{d_{n-1}}^{d_{n}}{\rm {Pr}}\left[\frac{P\zeta_{n}^{{\rm {NL}}}\left(r\right)h}{I_{{\rm {agg}}}+P_{{\rm {N}}}}>\gamma\right]f_{R,n}^{{\rm {NL}}}\left(r\right)dr$,
and $d_{0}$ and $d_{N}$ are defined as $0$ and $+\infty$, respectively.
Moreover, $f_{R,n}^{{\rm {L}}}\left(r\right)$ and $f_{R,n}^{{\rm {NL}}}\left(r\right)$
$\left(d_{n-1}<r\leq d_{n}\right)$, are represented by
\begin{equation}
f_{R,n}^{{\rm {L}}}\left(r\right)=\exp\left(\hspace{-0.1cm}-\hspace{-0.1cm}\int_{0}^{r_{1}}\left(1-{\rm {Pr}}^{{\rm {L}}}\left(u\right)\right)2\pi u\lambda du\right)\exp\left(\hspace{-0.1cm}-\hspace{-0.1cm}\int_{0}^{r}{\rm {Pr}}^{{\rm {L}}}\left(u\right)2\pi u\lambda du\right){\rm {Pr}}_{n}^{{\rm {L}}}\left(r\right)2\pi r\lambda,\hspace{-0.1cm}\hspace{-0.1cm}\hspace{-0.1cm}\hspace{-0.1cm}\label{eq:geom_dis_PDF_UAS1_LoS_thm}
\end{equation}
and
\begin{equation}
\hspace{-0.1cm}\hspace{-0.1cm}\hspace{-0.1cm}\hspace{-0.1cm}\hspace{-0.1cm}\hspace{-0.1cm}f_{R,n}^{{\rm {NL}}}\left(r\right)=\exp\left(\hspace{-0.1cm}-\hspace{-0.1cm}\int_{0}^{r_{2}}{\rm {Pr}}^{{\rm {L}}}\left(u\right)2\pi u\lambda du\right)\exp\left(\hspace{-0.1cm}-\hspace{-0.1cm}\int_{0}^{r}\left(1-{\rm {Pr}}^{{\rm {L}}}\left(u\right)\right)2\pi u\lambda du\right)\left(1-{\rm {Pr}}_{n}^{{\rm {L}}}\left(r\right)\right)2\pi r\lambda,\hspace{-0.1cm}\hspace{-0.1cm}\hspace{-0.1cm}\hspace{-0.1cm}\label{eq:geom_dis_PDF_UAS1_NLoS_thm}
\end{equation}
where $r_{1}$ and $r_{2}$ are given implicitly by the following
equations as
\begin{equation}
r_{1}=\underset{r_{1}}{\arg}\left\{ \zeta^{{\rm {NL}}}\left(r_{1}\right)=\zeta_{n}^{{\rm {L}}}\left(r\right)\right\} ,\label{eq:def_r_1}
\end{equation}
and
\begin{equation}
r_{2}=\underset{r_{2}}{\arg}\left\{ \zeta^{{\rm {L}}}\left(r_{2}\right)=\zeta_{n}^{{\rm {NL}}}\left(r\right)\right\} .\label{eq:def_r_2}
\end{equation}
In addition, ${\rm {Pr}}\left[\frac{P\zeta_{n}^{{\rm {L}}}\left(r\right)h}{I_{{\rm {agg}}}+P_{{\rm {N}}}}>\gamma\right]$
and ${\rm {Pr}}\left[\frac{P\zeta_{n}^{{\rm {NL}}}\left(r\right)h}{I_{{\rm {agg}}}+P_{{\rm {N}}}}>\gamma\right]$
are respectively computed by
\begin{equation}
{\rm {Pr}}\left[\frac{P\zeta_{n}^{{\rm {L}}}\left(r\right)h}{I_{{\rm {agg}}}+P_{{\rm {N}}}}>\gamma\right]=\exp\left(-\frac{\gamma P_{{\rm {N}}}}{P\zeta_{n}^{{\rm {L}}}\left(r\right)}\right)\mathscr{L}_{I_{{\rm {agg}}}}^{{\rm {L}}}\left(\frac{\gamma}{P\zeta_{n}^{{\rm {L}}}\left(r\right)}\right),\label{eq:Pr_SINR_req_UAS1_LoS_thm}
\end{equation}
where $\mathscr{L}_{I_{{\rm {agg}}}}^{{\rm {L}}}\left(s\right)$ is
the Laplace transform of $I_{{\rm {agg}}}$ for LoS signal transmission
evaluated at $s$, which can be further written as
\begin{equation}
\mathscr{L}_{I_{{\rm {agg}}}}^{{\rm {L}}}\left(s\right)=\exp\left(-2\pi\tilde{\lambda}\int_{r}^{+\infty}\frac{{\rm {Pr}}^{{\rm {L}}}\left(u\right)u}{1+\left(sP\zeta^{{\rm {L}}}\left(u\right)\right)^{-1}}du\right)\exp\left(-2\pi\tilde{\lambda}\int_{r_{1}}^{+\infty}\frac{\left[1-{\rm {Pr}}^{{\rm {L}}}\left(u\right)\right]u}{1+\left(sP\zeta^{{\rm {NL}}}\left(u\right)\right)^{-1}}du\right),\label{eq:laplace_term_LoS_UAS1_general_seg_thm}
\end{equation}
and
\begin{equation}
{\rm {Pr}}\left[\frac{P\zeta_{n}^{{\rm {NL}}}\left(r\right)h}{I_{{\rm {agg}}}+P_{{\rm {N}}}}>\gamma\right]=\exp\left(-\frac{\gamma P_{{\rm {N}}}}{P\zeta_{n}^{{\rm {NL}}}\left(r\right)}\right)\mathscr{L}_{I_{{\rm {agg}}}}^{{\rm {NL}}}\left(\frac{\gamma}{P\zeta_{n}^{{\rm {NL}}}\left(r\right)}\right),\label{eq:Pr_SINR_req_UAS1_NLoS_thm}
\end{equation}
where $\mathscr{L}_{I_{{\rm {agg}}}}^{{\rm {NL}}}\left(s\right)$
is the Laplace transform of $I_{{\rm {agg}}}$ for NLoS signal transmission
evaluated at $s$, which can be further written as
\begin{equation}
\mathscr{L}_{I_{{\rm {agg}}}}^{{\rm {NL}}}\left(s\right)=\exp\left(-2\pi\tilde{\lambda}\int_{r_{2}}^{+\infty}\frac{{\rm {Pr}}^{{\rm {L}}}\left(u\right)u}{1+\left(sP\zeta^{{\rm {L}}}\left(u\right)\right)^{-1}}du\right)\exp\left(-2\pi\tilde{\lambda}\int_{r}^{+\infty}\frac{\left[1-{\rm {Pr}}^{{\rm {L}}}\left(u\right)\right]u}{1+\left(sP\zeta^{{\rm {NL}}}\left(u\right)\right)^{-1}}du\right).\label{eq:laplace_term_NLoS_UAS1_general_seg_thm}
\end{equation}
\end{doublespace}
\end{thm}
\vspace{-1cm}

\begin{IEEEproof}
\begin{doublespace}
The proof is very similar to that for Theorem~1 in~\cite{our_work_TWC2016}.
Hence, we omit the proof here for brevity. The comparison between
Theorem~1 in~\cite{our_work_TWC2016} and the proposed theorem will
be explained in the sequel.
\end{doublespace}
\end{IEEEproof}
\end{algorithm*}
{\small \par}

From Theorem~\ref{thm:p_cov_UAS1} and comparing it with the main
result in~\cite{our_work_TWC2016}, which was derived for the case
with all BSs being active, it is important to note that:
\begin{itemize}
\item The impact of the serving BS selection on the coverage probability
is measured by (\ref{eq:geom_dis_PDF_UAS1_LoS_thm}) and (\ref{eq:geom_dis_PDF_UAS1_NLoS_thm}),
the expressions of which are based on $\lambda$, not on $\tilde{\lambda}$.
This is the same as Theorem~1 of~\cite{our_work_TWC2016}.
\item The impact of $I_{\textrm{agg}}$ on the coverage probability is measured
by (\ref{eq:laplace_term_LoS_UAS1_general_seg_thm}) and (\ref{eq:laplace_term_NLoS_UAS1_general_seg_thm}).
Since only the active BSs emit effective interference into the considered
SCN, the expressions of (\ref{eq:laplace_term_LoS_UAS1_general_seg_thm})
and (\ref{eq:laplace_term_NLoS_UAS1_general_seg_thm}) are thus based
on $\tilde{\lambda}$, not on $\lambda$. This is different from Theorem~1
of~\cite{our_work_TWC2016}.
\item The derivation of $\tilde{\lambda}$ is non-trivial, and it will be
treated later in the following subsections.
\end{itemize}

Besides, from Theorem~\ref{thm:p_cov_UAS1}, we can draw an important
intuition summarized in Lemma~\ref{lem:larger-CP}.
\begin{lem}
\noindent \textbf{\label{lem:larger-CP}}$p^{\textrm{cov}}\left(\lambda,\gamma\right)$
with the BS IMC is larger than that with all BSs being active.
\end{lem}
\begin{IEEEproof}
See Appendix~A.
\end{IEEEproof}

\subsection{The Area Spectral Efficiency\label{subsec:The-Area-Spectral}}

Similar to~\cite{our_work_TWC2016,our_GC_paper_2015_HPPP}, we also
investigate the area spectral efficiency (ASE) performance in $\textrm{bps/Hz/km}^{2}$,
which is defined as
\begin{equation}
A^{{\rm {ASE}}}\left(\lambda,\gamma_{0}\right)=\tilde{\lambda}\int_{\gamma_{0}}^{+\infty}\log_{2}\left(1+\gamma\right)f_{\mathit{\Gamma}}\left(\lambda,\gamma\right)d\gamma,\label{eq:ASE_def}
\end{equation}
where $\gamma_{0}$ is the minimum working SINR for the considered
SCN, and $f_{\mathit{\Gamma}}\left(\lambda,\gamma\right)$ is the
probability density function (PDF) of the SINR observed at the typical
UE at a particular value of $\lambda$. Based on the definition of
$p^{{\rm {cov}}}\left(\lambda,\gamma\right)$ in (\ref{eq:Coverage_Prob_def}),
which is the complementary cumulative distribution function (CCDF)
of SINR, $f_{\mathit{\Gamma}}\left(\lambda,\gamma\right)$ can be
computed by
\begin{equation}
f_{\mathit{\Gamma}}\left(\lambda,\gamma\right)=\frac{\partial\left(1-p^{{\rm {cov}}}\left(\lambda,\gamma\right)\right)}{\partial\gamma}.\label{eq:cond_SINR_PDF}
\end{equation}

Regarding $A^{{\rm {ASE}}}\left(\lambda,\gamma_{0}\right)$, it is
important to note that:
\begin{itemize}
\item Unlike~\cite{our_work_TWC2016,our_GC_paper_2015_HPPP}, in this work,
$\tilde{\lambda}$ is used in the expression of $A^{{\rm {ASE}}}\left(\lambda,\gamma_{0}\right)$
because only the active BSs make an effective contribution to the
ASE.
\item The ASE defined in this paper is different from that in~\cite{related_work_Jeff},
where a constant rate based on $\gamma_{0}$ is assumed for the typical
UE, no matter what the actual SINR value is. The definition of the
ASE in (\ref{eq:ASE_def}) can better capture the dependence of the
transmission rate on SINR, but it is less tractable to analyze, as
it requires one more fold of numerical integral compared with~\cite{related_work_Jeff}.
\item Previously in Subsection~\ref{subsec:The-Coverage-Probability},
we have obtained a conclusion from Theorem~\ref{thm:p_cov_UAS1}:
$p^{\textrm{cov}}\left(\lambda,\gamma\right)$ with the BS IMC should
be better than that with all BSs being active in dense SCNs due to
$\tilde{\lambda}\leq\lambda$. Here from (\ref{eq:ASE_def}), we may
arrive at an opposite conclusion for $A^{{\rm {ASE}}}\left(\lambda,\gamma_{0}\right)$.
The reasons are addressed as follows,
\begin{itemize}
\item In practice, there is a finite number of active UEs in the network,
and thus some BSs can be put to sleep in ultra-dense SCNs. As a result,
the spatial reuse factor of spectrum in an ultra-dense SCN is fundamentally
limited by the UE density $\rho$, and not by the BS density $\lambda$.
The extreme case happens where there is one UE per cell, thus there
cannot be more active BS than UEs.
\item However, if we assume that an infinite number of active UEs in the
network to activate all existing BSs, then the spatial reuse factor
of spectrum is then limited by the BS density $\lambda$.
\item In the former case, the inter-cell interference is severely bounded/mitigated
thanks to the less aggressive reuse factor of spectrum (i.e., in ultra-dense
SCNs, the UE density $\rho$ is relatively small compared with the
BS density $\lambda$, and thus many BS are put to sleep), which leads
to an enhanced performance per UE. However, the ASE is smaller than
that of the latter case. This is because less cells are active to
reuse the spectrum. Note that the ASE scales linearly with the spatial
reuse factor of spectrum. Thus, a head-to-head comparison of the ASE
with an infinite number of UEs and that with a finite number of UEs
is not fair.
\item To sum up, the takeaway message should not be that the IMC generates
an inferior ASE in dense SCNs. %
The key advantage of the BS IMC is that the per-UE performance should
increase with the network densification, which is a good performance
metric when considering a realistic finite number of UEs. %
\end{itemize}
\end{itemize}

\subsection{The Energy Efficiency\label{subsec:The-Energy-Efficiency}}

Deploying dense SCNs poses some concerns in terms of energy consumption.%
{} Hence, the energy efficiency (EE) of dense SCNs should be carefully
considered to allow for their sustainable deployments. When evaluating
the BS energy consumption, it is very important to note that a BS
in idle mode may still consume \emph{a non-negligible amount of energy},
thus impacting the EE of SCNs. In order to study realistic 5G networks,
here we use a practical power model developed in the Green-Touch project~\cite{DessetPowerModelling}.
This power model estimates the power consumption of a cellular BS,
and is based on tailored modeling principles and scaling rules for
each BS component i.e., power amplifier, analogue front-end, digital
base band, digital control and backhaul interface and power supply.
Moreover, it includes different optimized idle modes and provides
a large flexibility, i.e., multiple BS types are available, which
can be configured with multiple parameters, such as bandwidth, transmit
power, number of antenna chains, system load, duplex mode, etc. Among
the provided idle modes in the Green-Touch project, we consider \emph{the
Green-Touch slow idle mode} and \emph{the Green-Touch shut-down mode},
where most components of an idle BS are deactivated. Note that these
two modes are the most energy-efficient ones defined by the Green-Touch
project~\cite{DessetPowerModelling}.

Here, the total power of each idle SCN BS and that of each active
SCN BS are respectively denoted by $P_{{\rm {IMC}}}^{{\rm {TOT}}}\left(\lambda\right)$
and $P_{{\rm {ACT}}}^{{\rm {TOT}}}\left(\lambda\right)$, then we
can define the EE in the unit of $\textrm{bits/J}$ for the considered
SCN as
\begin{equation}
EE\left(\lambda,\gamma_{0}\right)=\frac{A^{{\rm {ASE}}}\left(\lambda,\gamma_{0}\right)\times BW}{\tilde{\lambda}P_{{\rm {ACT}}}^{{\rm {TOT}}}\left(\lambda\right)+\left(\lambda-\tilde{\lambda}\right)P_{{\rm {IMC}}}^{{\rm {TOT}}}\left(\lambda\right)},\label{eq:EE_def}
\end{equation}
where the area spectral efficiency $A^{{\rm {ASE}}}\left(\lambda,\gamma_{0}\right)$
is obtained from (\ref{eq:ASE_def}) and $BW$ denotes the system
bandwidth in Hz.

It is important to note that $EE\left(\lambda,\gamma_{0}\right)$
should depend on $\tilde{\lambda}$. More specifically, in the numerator
of $EE\left(\lambda,\gamma_{0}\right)$, we have $A^{{\rm {ASE}}}\left(\lambda,\gamma_{0}\right)$,
which scales linearly with respect to $\tilde{\lambda}$, as shown
in (\ref{eq:ASE_def}). Having said that, we would like to clarify
that $\tilde{\lambda}$ is a function of $\lambda$, as will be addressed
in the following Subsections. Therefore, we believe that $\lambda$
is a more fundamental variable than $\tilde{\lambda}$, and thus we
use $\lambda$ instead of $\tilde{\lambda}$ in $EE\left(\lambda,\gamma_{0}\right)$.

It is also important to note that in practice $P_{{\rm {IMC}}}^{{\rm {TOT}}}\left(\lambda\right)$
and $P_{{\rm {ACT}}}^{{\rm {TOT}}}\left(\lambda\right)$ should depend
on the BS density $\lambda$ because the BS transmission power decreases
with the network densification~\cite{Tutor_smallcell}. Nevertheless,
in previous subsections, we assume that the BS transmission power
$P$ is independent of $\lambda$ in (\ref{eq:SINR}) because (i)
it brings convenient expressions for our main results; and (ii) it
has a minor impact on the ASE performance for dense SCNs, since the
4G/5G network is interference limited and thus the BS transmission
power $P$ can be removed from both the numerator and the denominator
in the SINR expression (\ref{eq:SINR}).

From the results of $p^{{\rm {cov}}}\left(\lambda,\gamma\right)$,
$A^{{\rm {ASE}}}\left(\lambda,\gamma_{0}\right)$ and $EE\left(\lambda,\gamma_{0}\right)$,
respectively presented in~(\ref{eq:Coverage_Prob_def}),~(\ref{eq:ASE_def})
and~(\ref{eq:EE_def}), we can now analyze these performance measures
for the considered SCN. The key step to do so is to accurately derive
$\tilde{\lambda}$, i.e., the active BS density, which will be addressed
in the following subsections.

\subsection{A Lower Bound of $\tilde{\lambda}$\label{subsec:A-Lower-Bound}}

In~\cite{dynOnOff_Huang2012}, the authors derived an approximate
expression of $\tilde{\lambda}$ based on the distribution of the
Voronoi cell size assuming that each UE should be associated with
\emph{the nearest BS}. The main result in~\cite{dynOnOff_Huang2012}
is as follows,
\begin{equation}
\tilde{\lambda}^{{\rm {minDis}}}\approx\lambda\left[1-\frac{1}{\left(1+\frac{\rho}{q\lambda}\right)^{q}}\right]\overset{\triangle}{=}\lambda_{0}\left(q\right),\label{eq:lambda_tilde_Huang}
\end{equation}
where $\tilde{\lambda}^{{\rm {minDis}}}$ is the active BS density
under the assumption that each UE should connect to its nearest BS.
An empirical value of 3.5 was suggested for $q$ in~\cite{dynOnOff_Huang2012}.
The approximation was shown to be very accurate in existing work~\cite{dynOnOff_Huang2012,Li2014_energyEff,Zhang2016HetNetEE}
assuming a nearest-distance UAS. In this work, a more realistic signal
strength based UAS is adopted, and thus the corresponding result in~\cite{dynOnOff_Huang2012}
cannot be directly applied to Theorem~\ref{thm:p_cov_UAS1}. Instead,
we need to derive $\tilde{\lambda}$ for the adopted UAS considering
probabilistic LoS and NLoS transmissions, which will be addressed
step by step in the following subsections.

First, in Theorem~\ref{thm:lambda_tilde_LB}, we propose that $\tilde{\lambda}^{{\rm {minDis}}}$
in (\ref{eq:lambda_tilde_Huang}) is a lower bound of $\tilde{\lambda}$.
\begin{thm}
\label{thm:lambda_tilde_LB}Based on the path loss model in (\ref{eq:prop_PL_model})
and the presented UAS, $\tilde{\lambda}$ can be lower bounded by
\begin{equation}
\tilde{\lambda}\geq\tilde{\lambda}^{{\rm {minDis}}}\overset{\triangle}{=}\tilde{\lambda}^{{\rm {LB}}}.\label{eq:lambda_tilde_LB}
\end{equation}
\end{thm}
\begin{IEEEproof}
See Appendix~B.
\end{IEEEproof}

Intuitively speaking, the proof of Theorem~\ref{thm:lambda_tilde_LB}
states that from a typical UE's point of view, the equivalent BS density
of the considered UAS based on probabilistic LoS and NLoS transmissions
should be larger than that of the nearest-distance UAS based on single-slope
path loss transmissions. In other words, the existence of LoS BSs
provides more candidate BSs for a typical UE to connect with, and
thus the equivalent BS density increases for each UE. Since a larger
$\lambda$ always leads to a larger $\tilde{\lambda}$ due to a higher
BS diversity, we have $\tilde{\lambda}\geq\tilde{\lambda}^{{\rm {minDis}}}$.
As discussed before, the exact expression of $\tilde{\lambda}^{{\rm {minDis}}}$
is still unknown up to now, but it can be well approximated by $\lambda_{0}\left(q\right)$
shown in (\ref{eq:lambda_tilde_Huang}). The tightness of $\tilde{\lambda}^{{\rm {LB}}}$
will be verified using numerical results in Section~\ref{sec:Simulation-and-Discussion}.

\subsection{An Upper Bound of $\tilde{\lambda}$\label{subsec:An-Upper-Bound}}

Next, we propose an upper bound of $\tilde{\lambda}$ in Theorem~\ref{thm:lambda_tilde_UB}.
\begin{thm}
\label{thm:lambda_tilde_UB}Based on the path loss model in (\ref{eq:prop_PL_model})
and the presented UAS, $\tilde{\lambda}$ can be upper bounded by
\begin{equation}
\tilde{\lambda}\leq\lambda\left(1-Q^{{\rm {off}}}\right)\overset{\triangle}{=}\tilde{\lambda}^{{\rm {UB}}},\label{eq:lambda_tilde_UB_thm}
\end{equation}
where
\begin{equation}
Q^{{\rm {off}}}=\underset{r_{{\rm {max}}}\rightarrow+\infty}{\lim}\sum_{k=0}^{+\infty}\left\{ {\rm {Pr}}\left[w\nsim b\right]\right\} ^{k}\frac{\lambda_{\Omega}^{k}e^{-\lambda_{\Omega}}}{k!},\label{eq:Q_off_thm}
\end{equation}
where $\lambda_{\Omega}=\rho\pi r_{{\rm {max}}}^{2}$, and ${\rm {Pr}}\left[w\nsim b\right]$
represents the probability that a UE $w$ is not associated with BS
$b$ and it can be computed by
\begin{equation}
{\rm {Pr}}\left[w\nsim b\right]=\int_{0}^{r_{{\rm {max}}}}{\rm {Pr}}\left[\left.w\nsim b\right|r\right]\frac{2r}{r_{{\rm {max}}}^{2}}dr,\label{eq:Pr_w_notAss_b_thm}
\end{equation}
and
\begin{eqnarray}
\hspace{-0.2cm}\hspace{-0.2cm}\hspace{-0.2cm}{\rm {Pr}}\left[\left.w\nsim b\right|r\right]\hspace{-0.2cm} & = & \hspace{-0.2cm}\left[F_{R}^{{\rm {L}}}\left(r\right)+F_{R}^{{\rm {NL}}}\left(r_{1}\right)\right]{\rm {Pr}}^{{\rm {L}}}\left(r\right)\nonumber \\
\hspace{-0.2cm} &  & \hspace{-0.2cm}+\left[F_{R}^{{\rm {L}}}\left(r_{2}\right)+F_{R}^{{\rm {NL}}}\left(r\right)\right]\left[1-{\rm {Pr}}^{{\rm {L}}}\left(r\right)\right],\label{eq:Pr_w_notAss_b_cond_r_thm}
\end{eqnarray}
where $F_{R}^{{\rm {L}}}\left(r\right)=\int_{0}^{r}f_{R}^{{\rm {L}}}\left(u\right)du$,
$F_{R}^{{\rm {NL}}}\left(r\right)=\int_{0}^{r}f_{R}^{{\rm {NL}}}\left(u\right)du$,
and $r_{1}$ and $r_{2}$ are defined in (\ref{eq:def_r_1}) and (\ref{eq:def_r_2}),
respectively.
\end{thm}
\begin{IEEEproof}
See Appendix~C.
\end{IEEEproof}

Intuitively speaking, the proof of Theorem~\ref{thm:lambda_tilde_UB}
checks a disk area $\Omega$ centered on a typical BS (with a radius
of $r_{{\rm {max}}}$ and $r_{{\rm {max}}}\rightarrow+\infty$ in
(\ref{eq:Q_off_thm})), and calculate the probability that there is
no UE inside $\Omega$ connecting to this typical BS, i.e., the probability
$Q^{{\rm {off}}}$ that the typical BS should enter an idle mode.
In the computation of $Q^{{\rm {off}}}$, we ignore the serving BS
correlation between nearby UEs inside $\Omega$, i.e., the correlation
that a UE \emph{$k$ not} associated with BS $b$ may imply a nearby
UE $k'$ \emph{also not} associated with BS $b$ with a large probability.
This might be caused by another BS $b'$ located in the vicinity of
BS $b$. Therefore, here we under-estimate $Q^{{\rm {off}}}$, which
leads to an over-estimation of $\tilde{\lambda}$ as $\lambda\left(1-Q^{{\rm {off}}}\right)$
in (\ref{eq:lambda_tilde_UB_thm}). The tightness of $\tilde{\lambda}^{{\rm {UB}}}$
will be verified using numerical results in Section~\ref{sec:Simulation-and-Discussion}.

\subsection{The Proposed Approximation of $\tilde{\lambda}$\label{subsec:The-Proposed-Approximation}}

Considering the good tightness of the lower bound $\tilde{\lambda}^{{\rm {LB}}}$
to be shown in Section~\ref{sec:Simulation-and-Discussion}, and
the fact that the approximate expression of $\tilde{\lambda}^{{\rm {LB}}}$
is \emph{an increasing function} with respect to $q$, we propose
Proposition~\ref{prop:approx_lambda_tilde} to obtain an approximate
value of $\tilde{\lambda}$.
\begin{prop}
\label{prop:approx_lambda_tilde}Based on the path loss model in (\ref{eq:prop_PL_model})
and the adopted UAS, we propose to approximate $\tilde{\lambda}$
by
\begin{equation}
\tilde{\lambda}\approx\lambda_{0}\left(q^{*}\right),\label{eq:lambda_tilde_approx}
\end{equation}
where $3.5\leq q^{*}\leq\underset{x}{\arg}\left\{ \lambda_{0}\left(x\right)=\tilde{\lambda}^{{\rm {UB}}}\right\} $
and $\tilde{\lambda}^{{\rm {UB}}}$ is computed from (\ref{eq:lambda_tilde_UB_thm}).
\end{prop}

Note that the range of $q^{*}$ in Proposition~\ref{prop:approx_lambda_tilde}
is obtained according to the derived lower bound $\tilde{\lambda}^{{\rm {LB}}}$
and the upper bound $\tilde{\lambda}^{{\rm {UB}}}$ presented in Theorem~\ref{thm:lambda_tilde_LB}
and Theorem~\ref{thm:lambda_tilde_UB}, respectively. Apparently,
the value of $q^{*}$ depends on the specific forms of the path loss
model given by (\ref{eq:general_PL_func}) and (\ref{eq:general_LoS_Pr}).
Hence, $q^{*}$ should be numerically found for specific path loss
models in consideration. Fortunately, with the deterministic bounds
of $q^{*}$ characterized in Proposition~\ref{prop:approx_lambda_tilde},
the value of $q^{*}$ can be efficiently found using offline computation
based on the bisection method~\cite{Bisection} by minimizing the
difference between the approximate results of $\tilde{\lambda}$ in
(\ref{eq:lambda_tilde_approx}) and the simulated ones. Such difference
should be accounted and averaged over all possible values of $\lambda$
because $\lambda_{0}\left(q^{*}\right)$ also varies with $\lambda$.
The average difference can be measured by, e.g., the mean squared
error (MSE), giving rise to the search of $q^{*}$ based on the minimum
MSE (MMSE) criterion.

\subsection{The 3GPP Special Cases\label{subsec:The-3GPP-Special-Cases}}

As a special case to show our analytical results, following~\cite{our_work_TWC2016},
we consider a two-piece path loss and a linear LoS probability functions
defined by the 3GPP~\cite{TR36.828,SCM_pathloss_model}. Specifically,
we use the path loss function $\zeta\left(r\right)$, defined in the
3GPP as~\cite{TR36.828}
\begin{equation}
\zeta\left(r\right)=\begin{cases}
\begin{array}{l}
A^{{\rm {L}}}r^{-\alpha^{{\rm {L}}}},\\
A^{{\rm {NL}}}r^{-\alpha^{{\rm {NL}}}},
\end{array}\hspace{-0.3cm} & \begin{array}{l}
\textrm{\textrm{LoS:}~}\textrm{Pr}^{\textrm{L}}\left(r\right)\\
\textrm{\textrm{NLoS:}~}1-\textrm{Pr}^{\textrm{L}}\left(r\right)
\end{array}\end{cases},\label{eq:PL_BS2UE_2slopes}
\end{equation}
together with a linear LoS probability function of $\textrm{Pr}^{\textrm{L}}\left(r\right)$,
defined in the 3GPP as~\cite{SCM_pathloss_model}
\begin{equation}
\textrm{Pr}^{\textrm{L}}\left(r\right)=\begin{cases}
\begin{array}{l}
1-\frac{r}{d_{1}},\\
0,
\end{array}\hspace{-0.3cm} & \begin{array}{l}
0<r\leq d_{1}\\
r>d_{1}
\end{array}\end{cases},\label{eq:LoS_Prob_func_linear}
\end{equation}
where $d_{1}=300$\ m~\cite{SCM_pathloss_model}. Considering the
general path loss model presented in (\ref{eq:prop_PL_model}), the
combined path loss model presented in (\ref{eq:PL_BS2UE_2slopes})
and (\ref{eq:LoS_Prob_func_linear}) can be deemed as a special case
of (\ref{eq:prop_PL_model}) with the following substitution: $N=2$,
$\zeta_{1}^{\textrm{L}}\left(r\right)=\zeta_{2}^{\textrm{L}}\left(r\right)=A^{{\rm {L}}}r^{-\alpha^{{\rm {L}}}}$,
$\zeta_{1}^{\textrm{NL}}\left(r\right)=\zeta_{2}^{\textrm{NL}}\left(r\right)=A^{{\rm {NL}}}r^{-\alpha^{{\rm {NL}}}}$,
$\textrm{Pr}_{1}^{\textrm{L}}\left(r\right)=1-\frac{r}{d_{1}}$, and
$\textrm{Pr}_{2}^{\textrm{L}}\left(r\right)=0$. For clarity, this
3GPP special case is referred to as \textbf{3GPP Case~1} in the sequel.
As justified in~\cite{our_work_TWC2016}, we mainly use 3GPP Case~1
to generate the numerical results in Section~\ref{sec:Simulation-and-Discussion},
because it provides tractable results for $\left\{ f_{R,n}^{Path}\left(r\right)\right\} $
and $\left\{ \mathscr{L}_{I_{{\rm {agg}}}}^{Path}\left(s\right)\right\} $
in (\ref{eq:geom_dis_PDF_UAS1_LoS_thm})-(\ref{eq:laplace_term_NLoS_UAS1_general_seg_thm})
of Theorem~\ref{thm:p_cov_UAS1}. %

Moreover, as another application of our analytical work and to demonstrate
that our conclusions have general significance, we consider another
widely used LoS probability function, which is a two-piece exponential
function defined in the 3GPP as~\cite{TR36.828,our_work_TWC2016}
\begin{equation}
\textrm{Pr}^{\textrm{L}}\left(r\right)=\begin{cases}
\begin{array}{l}
1-5\exp\left(-R_{1}/r\right),\\
5\exp\left(-r/R_{2}\right),
\end{array} & \begin{array}{l}
0<r\leq d_{1}\\
r>d_{1}
\end{array}\end{cases},\label{eq:LoS_Prob_func_reverseS_shape}
\end{equation}
where $R_{1}=156$\ m, $R_{2}=30$\ m, and $d_{1}=\frac{R_{1}}{\ln10}$~\cite{TR36.828}.%
{} For clarity, this combined case with both the path loss function
and the LoS probability function coming from~\cite{TR36.828} is
referred to as \textbf{3GPP Case~2} hereafter. Moreover, to make
3GPP Case~2 more practical than 3GPP Case~1, we further consider
\emph{distance-dependent Rician fading}~\cite{SCM_pathloss_model}
and\emph{ correlated shadow fading}~\cite{TR36.828} in 3GPP Case~2.
The details can be found in the last paragraph of Section~\ref{sec:System-Model}.
Due to the great difficulty in obtaining the analytical results for
3GPP Case~2, we will investigate 3GPP Case~2 using simulation in
Section~\ref{sec:Simulation-and-Discussion}, and show that similar
conclusions like those for 3GPP Case~1 can also be drawn for 3GPP
Case~2.

\section{Simulation and Discussion\label{sec:Simulation-and-Discussion}}

In this section, we investigate network performance and use numerical
results to validate the accuracy of our analysis. According to Tables
A.1-3, A.1-4 and A.1-7 of~\cite{TR36.828} and~\cite{SCM_pathloss_model},
we adopt the following parameters for 3GPP Case~1: $\alpha^{{\rm {L}}}=2.09$,
$\alpha^{{\rm {NL}}}=3.75$, $A^{{\rm {L}}}=10^{-10.38}$, $A^{{\rm {NL}}}=10^{-14.54}$%
, $BW=10$\ MHz, $P=24$\ dBm, $P_{{\rm {N}}}=-95$\ dBm (including
a noise figure of 9\ dB at each UE). Besides, the UE density $\rho$
is set to $100\thinspace\textrm{UEs/km}^{2}$, $300\thinspace\textrm{UEs/km}^{2}$
and $600\thinspace\textrm{UEs/km}^{2}$ to represent a SCN with a
low, medium and high traffic load, respectively~\cite{Tutor_smallcell}.

To evaluate the impact of different path loss models on our conclusions,
we have also investigated the results for a single-slope path loss
model that does not differentiate LoS and NLoS transmissions~\cite{Jeff2011}.
In such path loss model, one path loss exponent $\alpha$ is defined,
the value of which is assumed to be $\alpha=\alpha^{{\rm {NL}}}=3.75$.
Note that in this single-slope path loss model, the active BS density
is assumed to be $\lambda_{0}\left(3.5\right)$, shown in (\ref{eq:lambda_tilde_Huang})~\cite{dynOnOff_Huang2012}.%

\subsection{The Results of $\tilde{\lambda}$ for 3GPP Case~1\label{subsec:discussion_q_star_3GPPcase1}}

\begin{figure}[H]
\noindent \begin{centering}
\includegraphics[width=8cm]{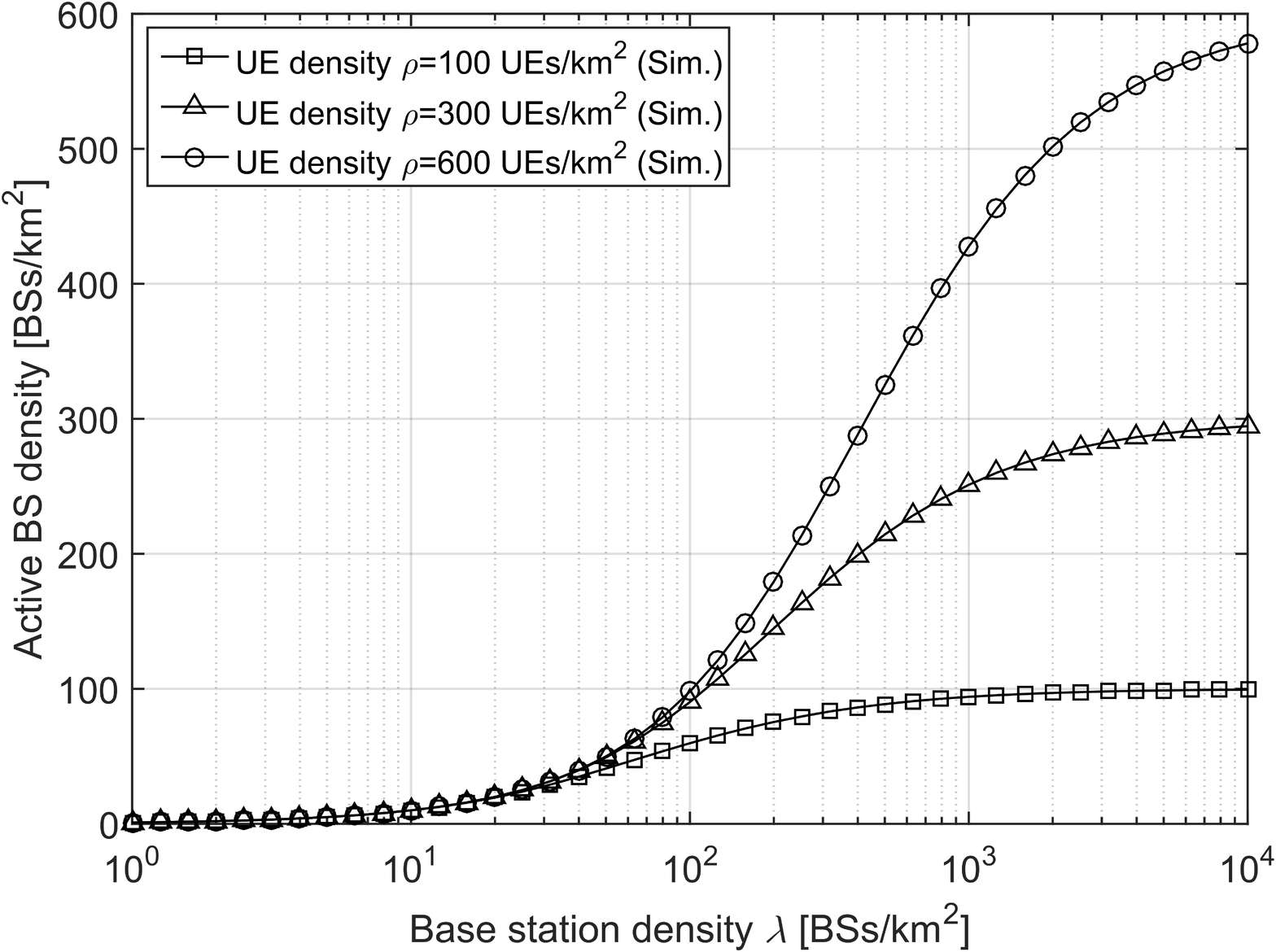}\renewcommand{\figurename}{Fig.}\caption{\label{fig:lambda_tilde_3GPPcase1_various_UEdensity}The active BS
density $\tilde{\lambda}$ with various values of $\rho$ for 3GPP
Case~1.}
\par\end{centering}
\vspace{-0.3cm}
\end{figure}

For 3GPP Case~1, the simulated results on the active BS density,
i.e., $\tilde{\lambda}$, for various values of $\rho$ are shown
in Fig.~\ref{fig:lambda_tilde_3GPPcase1_various_UEdensity}. As can
be seen from Fig.~\ref{fig:lambda_tilde_3GPPcase1_various_UEdensity},
more BSs will be activated with the network densification. However,
the value of $\tilde{\lambda}$ caps at $\rho$, because one UE can
activate at most one BS for its service.

\begin{figure}
\noindent \begin{centering}
\includegraphics[width=8cm]{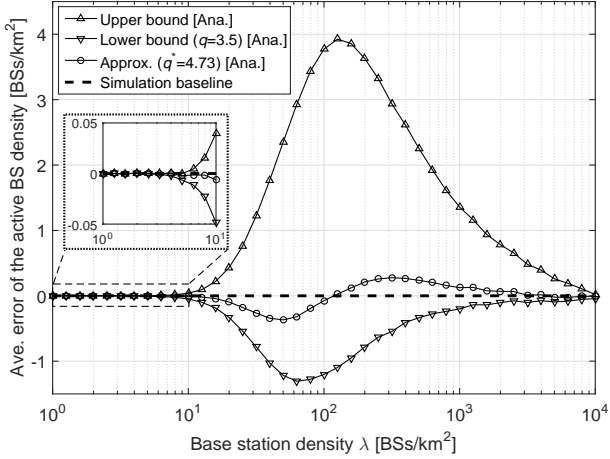}\renewcommand{\figurename}{Fig.}\caption{\label{fig:Average_error_actBSnum_UEdens100}The average error of
the active BS density for 3GPP Case~1 ($\rho=100\,\textrm{UEs/km}^{2}$).}
\par\end{centering}
\vspace{-0.3cm}
\end{figure}

\begin{figure}
\noindent \begin{centering}
\includegraphics[width=8cm]{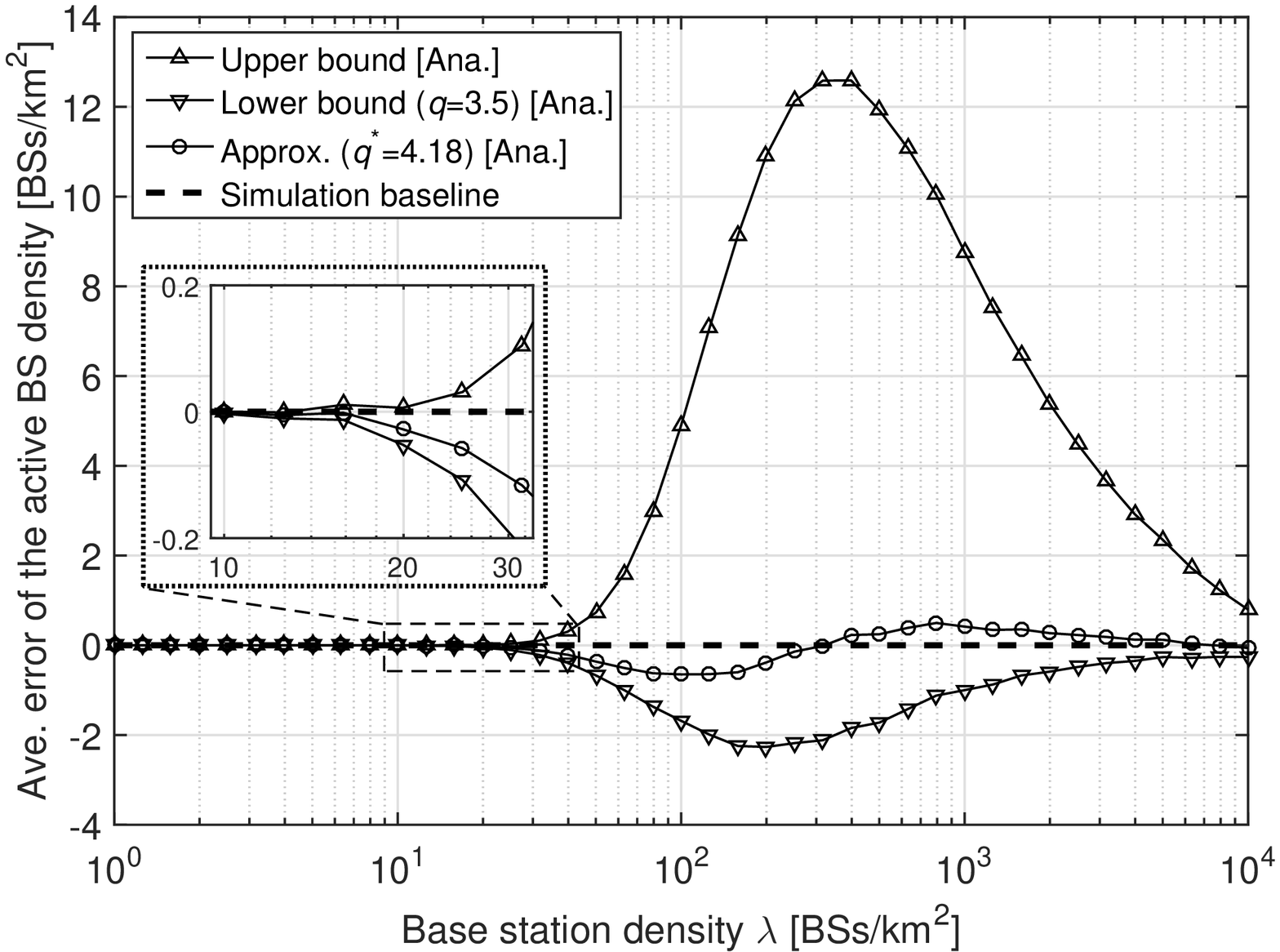}\renewcommand{\figurename}{Fig.}\caption{\label{fig:Average_error_actBSnum_UEdens300}Average error of the
active BS density for 3GPP Case~1 ($\rho=300\,\textrm{UEs/km}^{2}$).}
\par\end{centering}
\vspace{-0.3cm}
\end{figure}

\begin{figure}
\noindent \begin{centering}
\includegraphics[width=8cm]{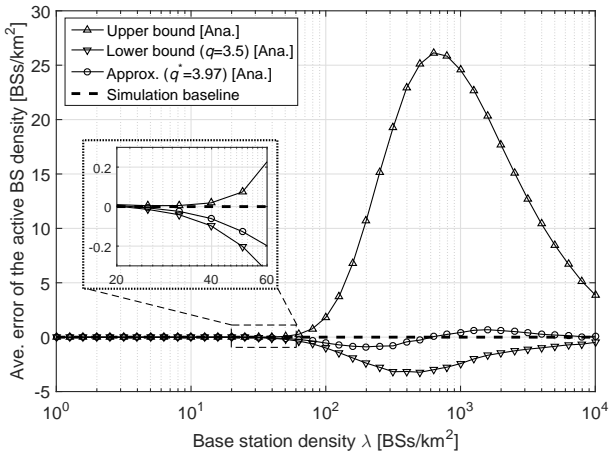}\renewcommand{\figurename}{Fig.}\caption{\label{fig:Average_error_actBSnum_UEdens600}Average error of the
active BS density for 3GPP Case~1 ($\rho=600\,\textrm{UEs/km}^{2}$).}
\par\end{centering}
\vspace{-0.3cm}
\end{figure}

Considering Proposition~\ref{prop:approx_lambda_tilde}, we conduct
a bisection search to numerically find the optimal $q^{*}$ for the
approximate $\tilde{\lambda}$. Based on the MMSE criterion proposed
in Subsection~\ref{subsec:The-Proposed-Approximation}, we obtain
$q^{*}=4.73$, $q^{*}=4.18$ and $q^{*}=3.97$ for the cases of $\rho=100\thinspace\textrm{UEs/km}^{2}$,
$\rho=300\thinspace\textrm{UEs/km}^{2}$ and $\rho=600\thinspace\textrm{UEs/km}^{2}$,
respectively.%
{} In Figs.~\ref{fig:Average_error_actBSnum_UEdens100},~\ref{fig:Average_error_actBSnum_UEdens300}
and~\ref{fig:Average_error_actBSnum_UEdens600}, we show the average
errors on the estimated values of $\tilde{\lambda}$ based on $\tilde{\lambda}^{{\rm {UB}}}$,
$\tilde{\lambda}^{{\rm {LB}}}$, and $\lambda_{0}\left(q^{*}\right)$.
\emph{Note that in these figures, all results are compared against
the simulation results shown in Fig.~\ref{fig:lambda_tilde_3GPPcase1_various_UEdensity},
which form the baseline results with zero errors. }Also note that
as discussed in Subsection~\ref{subsec:The-Area-Spectral}, the exact
expression of $\tilde{\lambda}^{{\rm {LB}}}$ is still unknown up
to now, but it can be well approximated by $\lambda_{0}\left(3.5\right)$,
presented in (\ref{eq:lambda_tilde_Huang}). Hence, the results of
$\lambda_{0}\left(3.5\right)$ are displayed in Fig.~\ref{fig:Average_error_actBSnum_UEdens300}
to represent an lower bound of $\tilde{\lambda}$.

As an example, from Fig.~\ref{fig:Average_error_actBSnum_UEdens300}
for $\rho=300\thinspace\textrm{UEs/km}^{2}$, we can draw the following
conclusions:
\begin{itemize}
\item The proposed upper bound $\tilde{\lambda}^{{\rm {UB}}}$ and lower
bound $\tilde{\lambda}^{{\rm {LB}}}$ are valid according to the simulation
results. More specifically, $\tilde{\lambda}^{{\rm {UB}}}$ and $\tilde{\lambda}^{{\rm {LB}}}$
are always larger (showing positive errors) and smaller (showing negative
errors) than the simulation baseline results, respectively.
\item $\tilde{\lambda}^{{\rm {UB}}}$ is tighter than $\tilde{\lambda}^{{\rm {LB}}}$
when $\lambda$ is relatively small, e.g., when $\lambda<30\,\textrm{BSs/km}^{2}$.%
\item $\tilde{\lambda}^{{\rm {LB}}}$ is much tighter than $\tilde{\lambda}^{{\rm {UB}}}$
for dense and ultra-dense SCNs, e.g., $\lambda>100\,\textrm{BSs/km}^{2}$.%
\item The maximum error associated with $\lambda_{0}\left(q^{*}\right)$
is smaller than those of $\tilde{\lambda}^{{\rm {UB}}}$ and $\tilde{\lambda}^{{\rm {LB}}}$,
e.g., when $\rho=300\thinspace\textrm{UEs/km}^{2}$ and $q^{*}=4.18$,
the maximum error resulting from $\lambda_{0}\left(q^{*}\right)$
is around $\pm$0.5\,$\textrm{BSs/km}^{2}$, while those given by
$\tilde{\lambda}^{{\rm {UB}}}$ and $\tilde{\lambda}^{{\rm {LB}}}$
are around 12\,$\textrm{BSs/km}^{2}$ and -2\,$\textrm{BSs/km}^{2}$,
respectively. Hence, $\lambda_{0}\left(q^{*}\right)$ gives a better
estimation on $\tilde{\lambda}$ than both $\tilde{\lambda}^{{\rm {UB}}}$
and $\tilde{\lambda}^{{\rm {LB}}}$.
\end{itemize}

\subsection{Validation of Theorem~\ref{thm:p_cov_UAS1} for 3GPP Case~1\label{subsec:Sim-p-cov-3GPP-Case-1}}

\begin{figure}
\noindent \begin{centering}
\includegraphics[width=8cm]{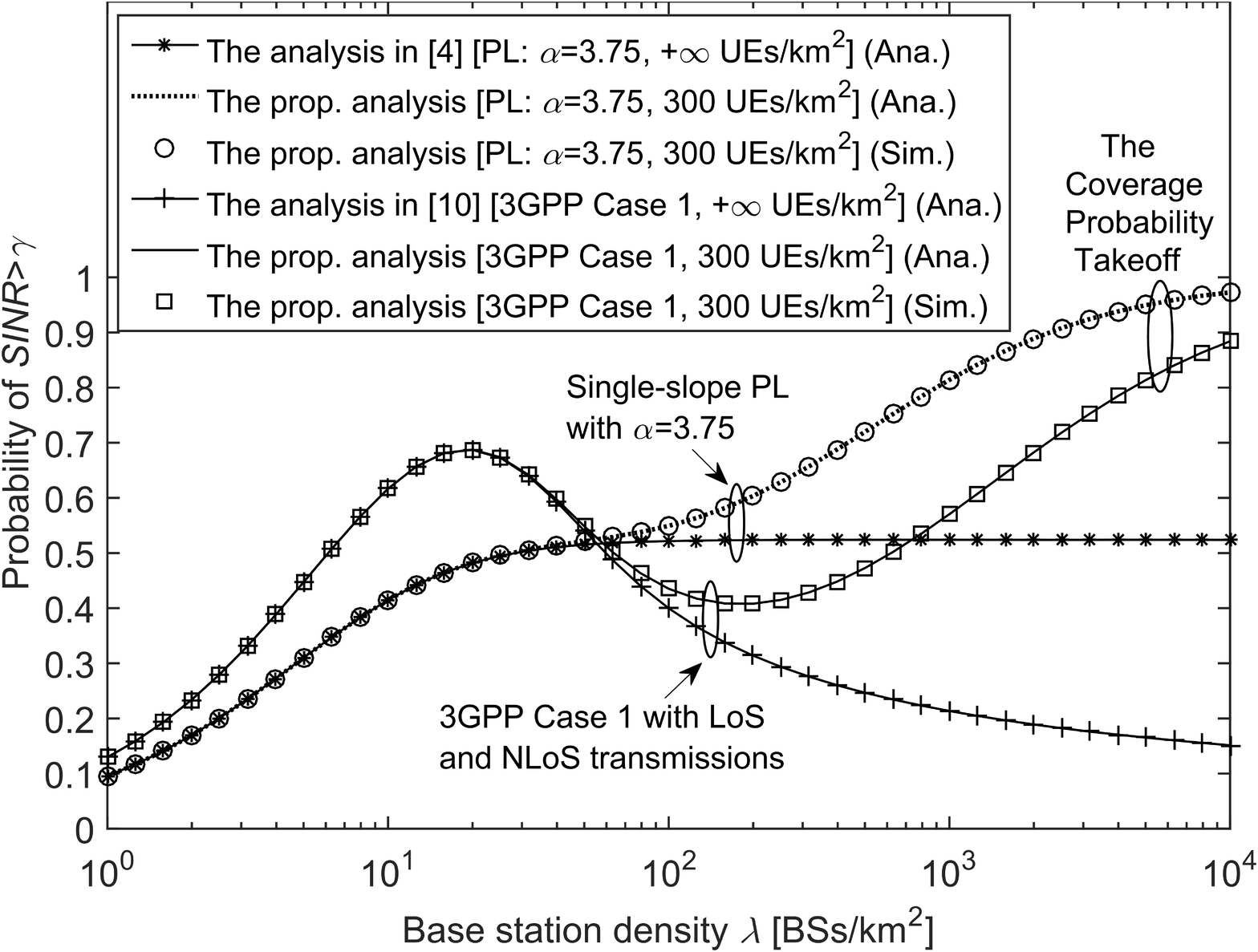}\renewcommand{\figurename}{Fig.}\caption{\label{fig:p_cov_vs_lambda_gamma0dB_UEdensity300}The coverage probability
$p^{\textrm{cov}}\left(\lambda,\gamma\right)$ vs. $\lambda$ for
3GPP Case~1 ($\gamma=0\,\textrm{dB}$, $\rho=300\,\textrm{UEs/km}^{2}$
and $q^{*}=4.18$). }
\par\end{centering}
\vspace{-0.3cm}
\end{figure}

In Fig.~\ref{fig:p_cov_vs_lambda_gamma0dB_UEdensity300}, we show
the results of $p^{\textrm{cov}}\left(\lambda,\gamma\right)$ when
$\rho=300\,\textrm{UEs/km}^{2}$ and $\gamma=0\,\textrm{dB}$, with
$q^{*}=4.18$ plugged into Proposition~\ref{prop:approx_lambda_tilde}.
As discussed in Section~\ref{sec:System-Model}, $\rho=300\,\textrm{UEs/km}^{2}$
is a typical density of active UEs in 5G~\cite{Tutor_smallcell},
which will be used to evaluate network performance in the following
subsections. Note that in our numerical results here and in the following
subsections, the proposed analysis is given by Theorem~\ref{thm:p_cov_UAS1}
and Proposition~\ref{prop:approx_lambda_tilde} with $q^{*}=4.18$.
As a benchmark, we also display the results for $\rho=+\infty\,\textrm{UEs/km}^{2}$
with all BSs being active.

As one can observe, our analytical results well match the simulation
results, which validates the accuracy of our analysis. In fact, Fig.~\ref{fig:p_cov_vs_lambda_gamma0dB_UEdensity300}
is essentially the same as Fig.~\ref{fig:comp_p_cov_4Gto5G}, except
that the results for the single-slope path loss model with $\rho=300\,\textrm{UEs/km}^{2}$
are also plotted here for a complete view of the performance behavior.
Moreover, Fig.~\ref{fig:p_cov_vs_lambda_gamma0dB_UEdensity300} confirms
the key observations presented in Section~\ref{sec:Introduction}:
\begin{itemize}
\item For the single-slope path loss model with $\rho=+\infty\,\textrm{UEs/km}^{2}$,
the coverage probability approaches a constant for dense SCNs, as
reported in~\cite{Jeff2011}. As $\rho$ approaches infinity, all
BSs are active. Thus, this scenario corresponds to a network condition
that does not require the IMC, i.e., the fully loaded network.
\item For 3GPP Case~1 with $\rho=+\infty\,\textrm{UEs/km}^{2}$, and when
the network is dense enough, i.e., $\lambda>20\,\textrm{BSs/km}^{2}$,
the coverage probability decreases as $\lambda$ increases due to
the NLoS to LoS transition of interference paths~\cite{our_work_TWC2016},
leading to a faster increase of the interference power compared with
the signal power.%
{}
\item For both path loss models with $\rho=300\,\textrm{UEs/km}^{2}$, the
coverage probability performance continuously increases toward one,
i.e., \emph{the Coverage Probability Takeoff.} This shows the benefits
of the IMC in dense SCNs, as discussed in Sections~\ref{sec:Introduction}
and~\ref{sec:Main-Results}.
\end{itemize}

\subsection{The ASE Performance for 3GPP Case~1\label{subsec:Sim-ASE-3GPP-Case-1}}

\begin{figure}[H]
\noindent \begin{centering}
\includegraphics[width=8cm]{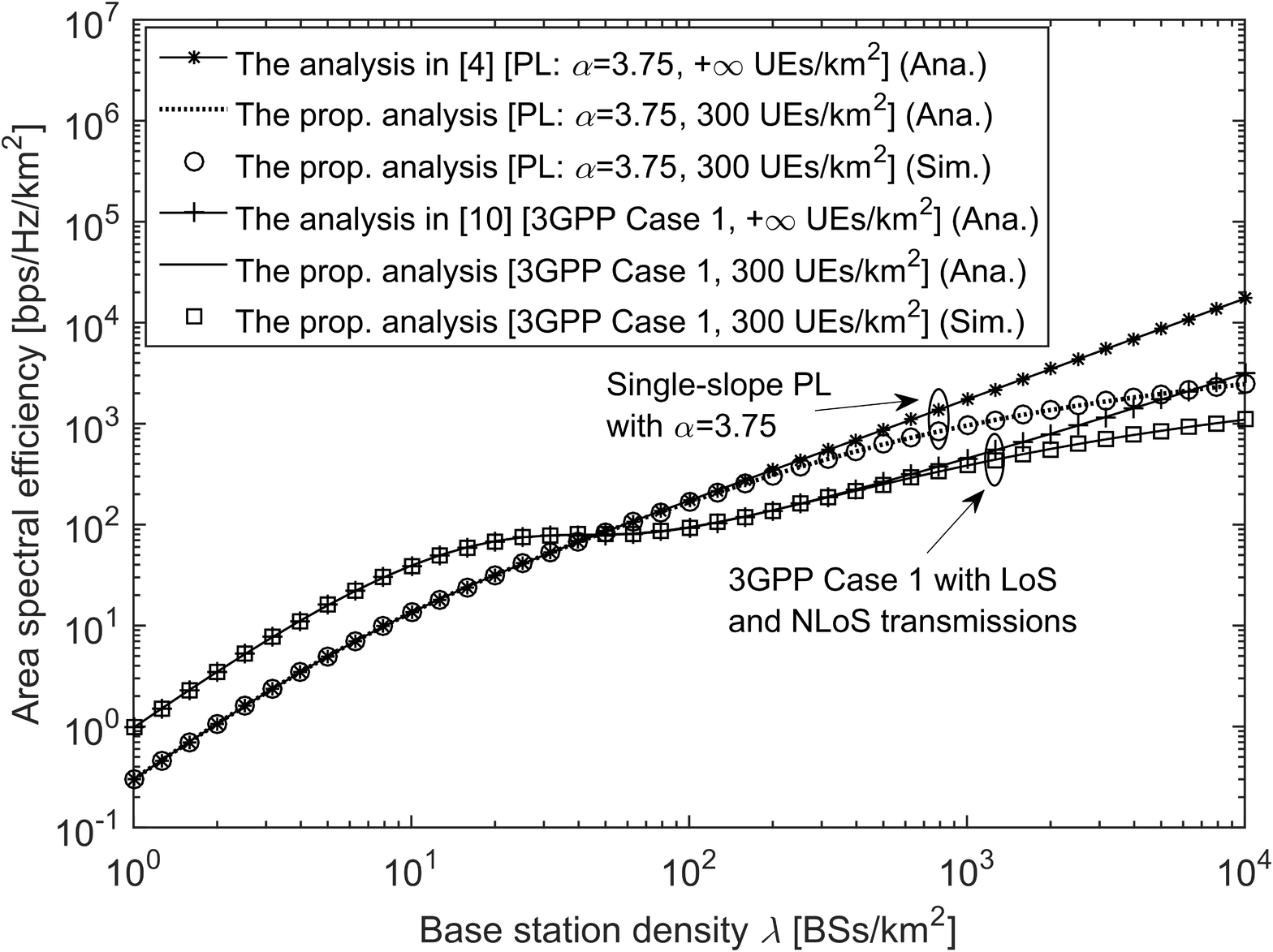}\renewcommand{\figurename}{Fig.}\caption{\label{fig:ASE_vs_lambda_gamma0dB_UEdensity300_fixedP}The ASE $A^{\textrm{ASE}}\left(\lambda,\gamma_{0}\right)$
vs. $\lambda$ for 3GPP Case~1 ($\gamma_{0}=0\,\textrm{dB}$, $\rho=300\,\textrm{UEs/km}^{2}$
and $q^{*}=4.18$).}
\par\end{centering}
\vspace{-0.3cm}
\end{figure}

In Fig.~\ref{fig:ASE_vs_lambda_gamma0dB_UEdensity300_fixedP}, we
plot the results of $A^{\textrm{ASE}}\left(\lambda,\gamma_{0}\right)$
when $\rho=300\,\textrm{UEs/km}^{2}$ and $\gamma_{0}=0\,\textrm{dB}$,
with $q^{*}=4.18$ plugged into Proposition~\ref{prop:approx_lambda_tilde}.
 From Fig.~\ref{fig:ASE_vs_lambda_gamma0dB_UEdensity300_fixedP},
we can draw the following conclusions:
\begin{itemize}
\item For 3GPP Case~1, the ASE suffers from a slow growth or even a slight
decrease when $\lambda\in\left[20,200\right]\,\textrm{BSs/km}^{2}$
because of the interference transition from NLoS to LoS~\cite{our_work_TWC2016}.
Such performance degradation has also been confirmed in Fig.~\ref{fig:p_cov_vs_lambda_gamma0dB_UEdensity300}.
\item After such BS density region of interference transition, for both
path loss models with $\rho=300\,\textrm{UEs/km}^{2}$ and the BS
IMC, the ASEs monotonically grow as $\lambda$ increases in dense
SCNs, but with noticeable performance gaps compared with those with
$\rho=+\infty\,\textrm{UEs/km}^{2}$.
\item As discussed in Section~\ref{sec:Main-Results}, the takeaway message
should not be that the IMC generates an inferior ASE in dense SCNs.
Instead, since there is a finite number of the active UEs in the network,
some BSs are put to sleep and thus the spatial spectrum reuse in practice
is fundamentally limited by $\rho$. The key advantage of the BS IMC
is that the per-UE performance should increase with the network densification
as exhibited in Fig.~\ref{fig:p_cov_vs_lambda_gamma0dB_UEdensity300}.
\end{itemize}

\subsection{The Performance of 3GPP Case~2\label{subsec:perfm_3GPP_Case2}}

\begin{figure*}
\center
\subfigure[The coverage probability $p^{\textrm{cov}}\left(\lambda,\gamma\right)$ vs. $\lambda$.]
{\includegraphics[width=7cm]{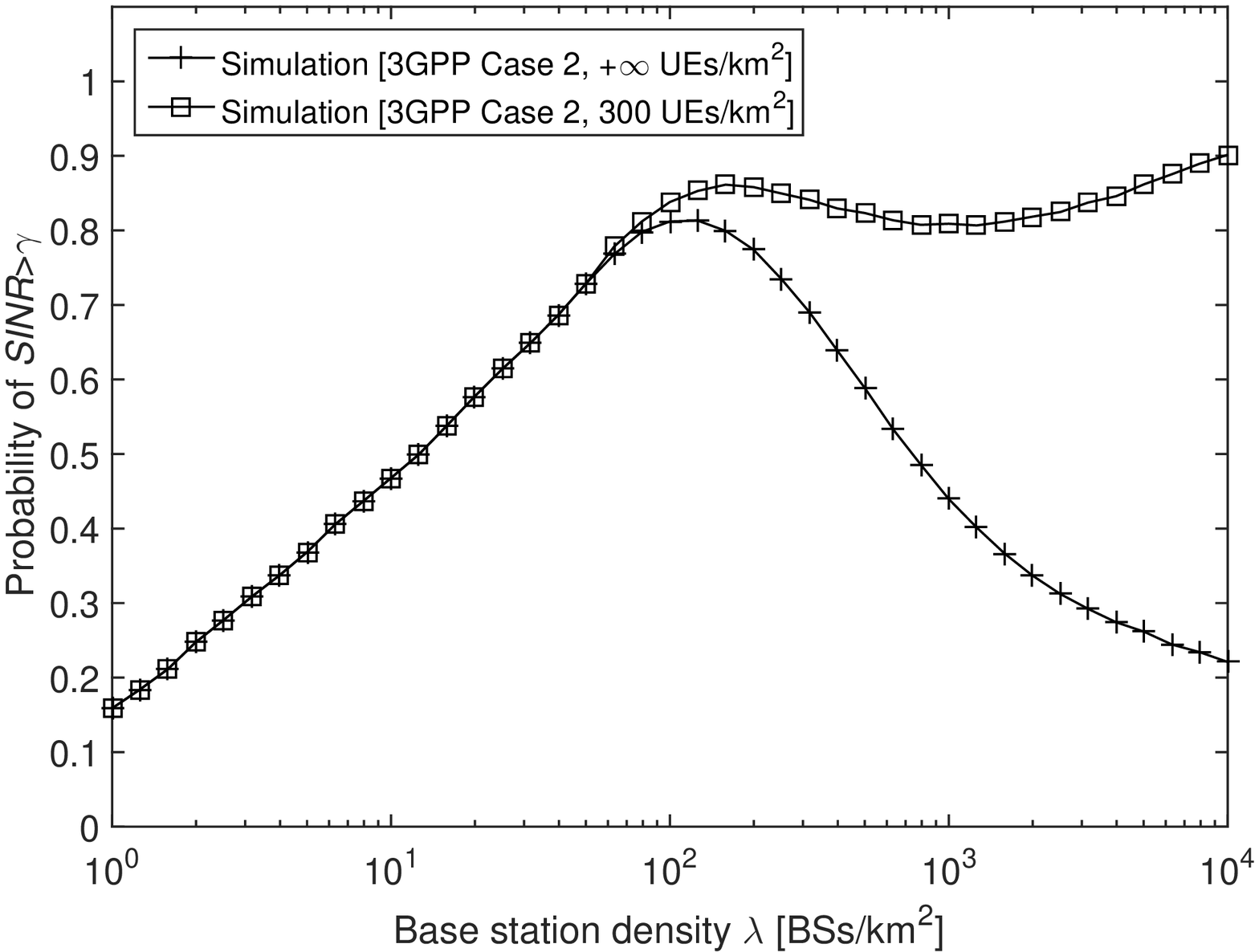}
\label{fig:p_cov_3GPP_case2}}
\hspace{0.5cm}
\subfigure[The ASE $A^{\textrm{ASE}}\left(\lambda,\gamma_{0}\right)$ vs. $\lambda$. ]{\includegraphics[width=7cm]{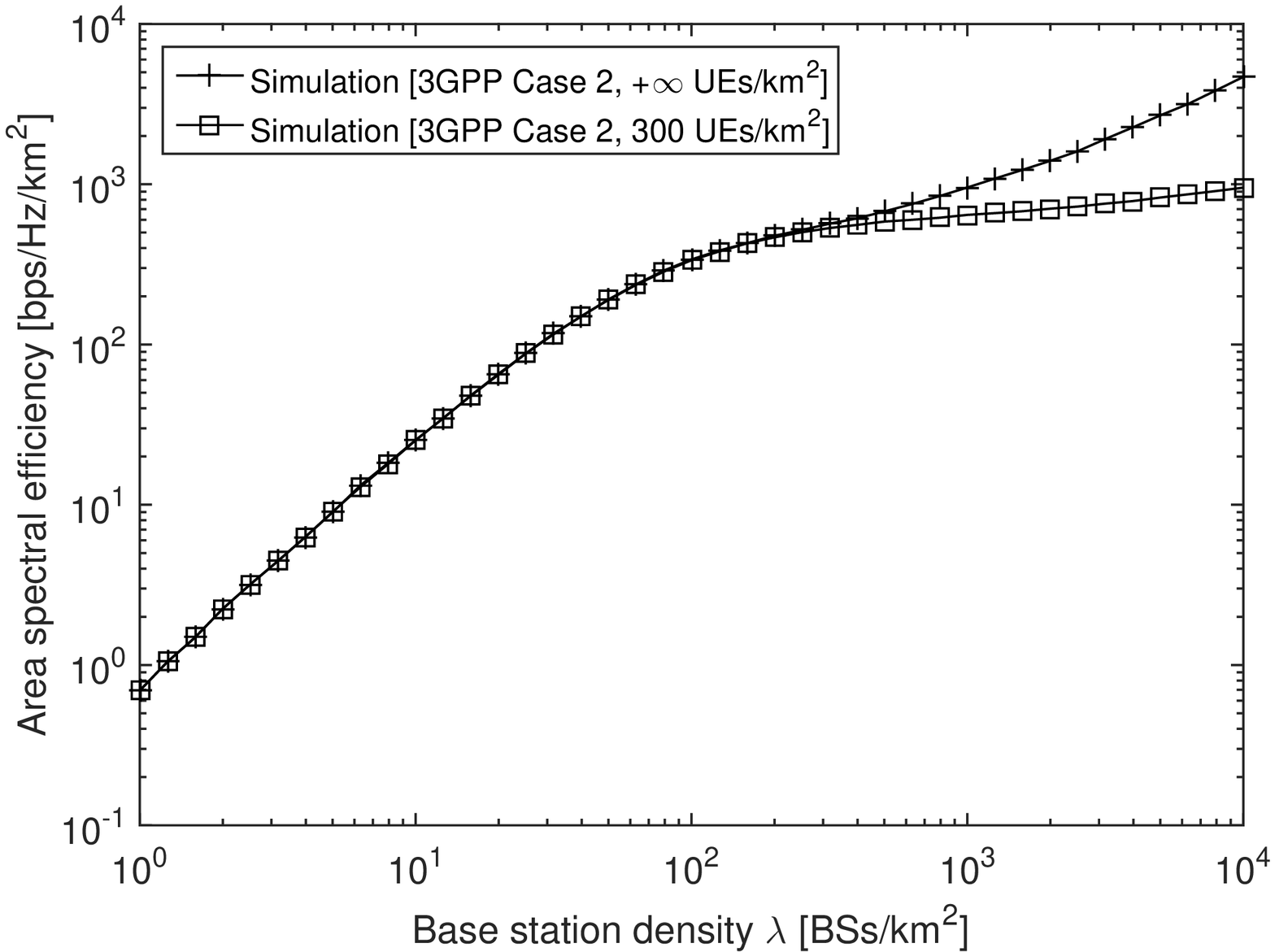}\label{fig:ASE_3GPP_case2}}
\caption{Performance for 3GPP Case 2 ($\gamma=0\,\textrm{dB}$ and $\rho=300\,\textrm{UEs/km}^{2}$).}
\label{fig:perfm_3GPP_case2}
\vspace{-0.3cm}
\end{figure*}

In this subsection, we investigate the performance for 3GPP Case~2
with an alternative path loss model, \emph{Rician fading} and \emph{correlated
shadow fading}, which have been discussed in Subsection~\ref{subsec:The-3GPP-Special-Cases}.
Due to the complex modeling of 3GPP Case~2, it is difficult to obtain
the analytical results for 3GPP Case~2. Hence, we conduct simulation
to investigate 3GPP Case~2 and the results are plotted in Fig.~\ref{fig:perfm_3GPP_case2}.
As one can observe from Fig.~\ref{fig:perfm_3GPP_case2}, all the
conclusions in Subsections~\ref{subsec:Sim-p-cov-3GPP-Case-1} and~\ref{subsec:Sim-ASE-3GPP-Case-1}
are qualitatively valid for Fig.~\ref{fig:perfm_3GPP_case2}. Only
some quantitative deviations exists, which shows the usefulness of
our theoretical analysis to predict the performance trend for dense
SCNs with the BS IMC.

\noindent 

\noindent \begin{figure*}
\center
\subfigure[The BS density dependent transmission power in dBm.]{\includegraphics[width=7cm]{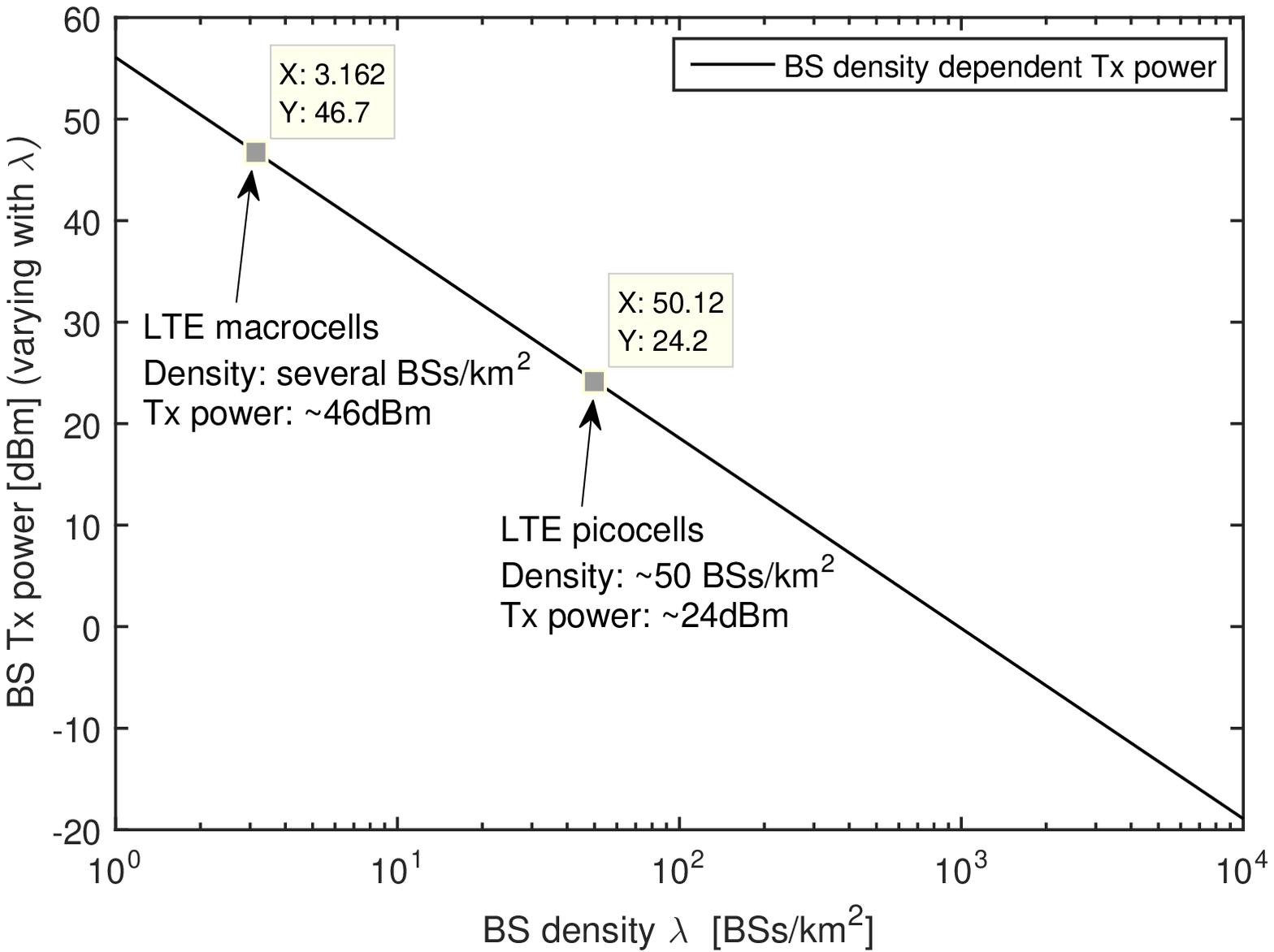}\label{fig:BS_den_depend_TxP}}
\hspace{0.5cm}
\subfigure[The BS density dependent total power in dBm.]{\includegraphics[width=7cm]{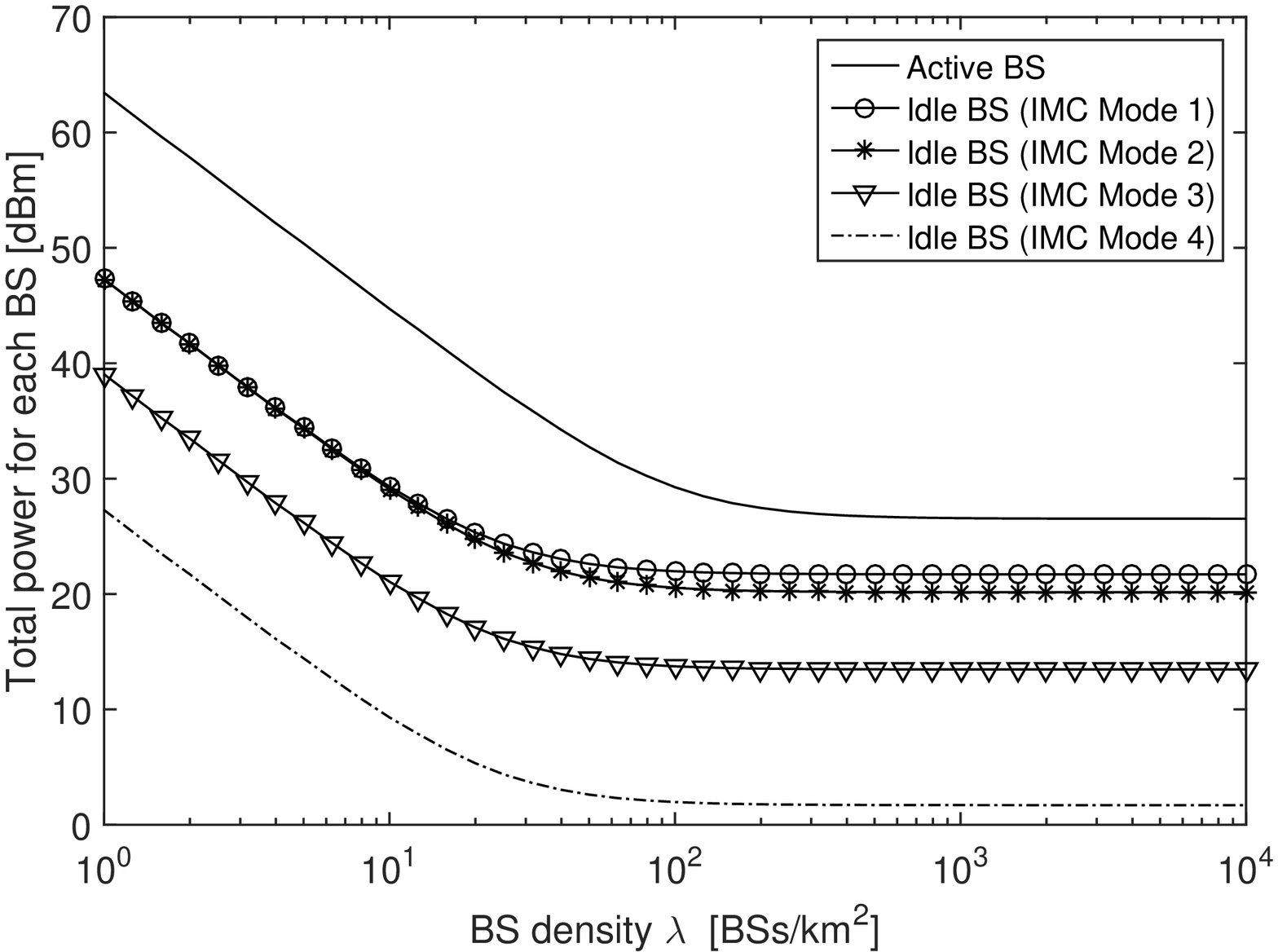}\label{fig:BS_den_depend_totP}}
\caption{The BS density dependent power configuration.}
\label{fig:BS_den_depend_P_config}
\vspace{-0.3cm}
\end{figure*}

\noindent \begin{figure*}
\center
\subfigure[The EE $EE\left(\lambda,\gamma_{0}\right)$ vs. $\lambda $ for 3GPP Case 1.]{\includegraphics[width=7cm]{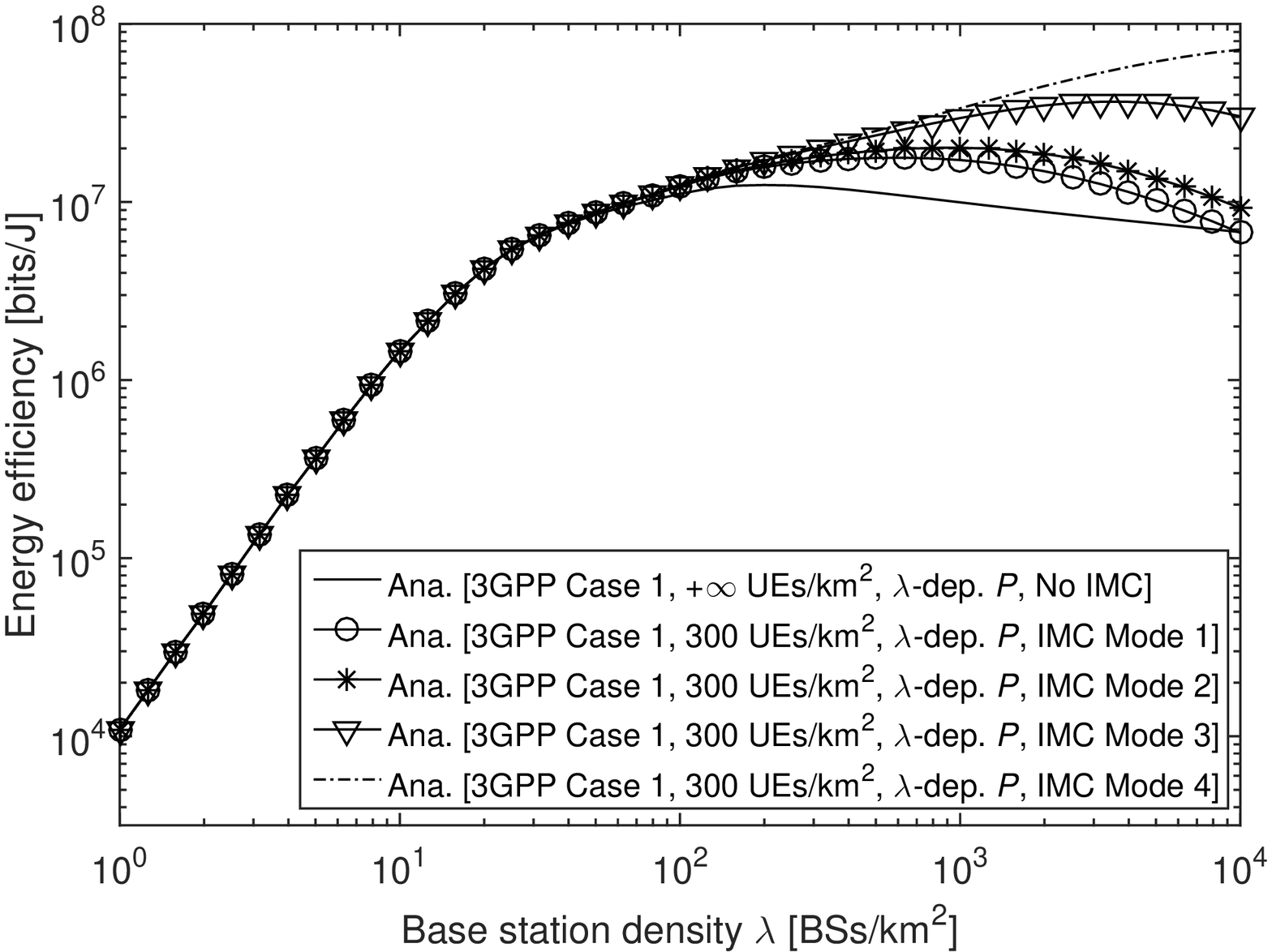}
\label{fig:EE_vs_lambda_gamma0dB_UEdensity300_varP_3GPP_case1}}
\hspace{0.5cm}
\subfigure[The EE $EE\left(\lambda,\gamma_{0}\right)$ vs. $\lambda $ for 3GPP Case 2.]{\includegraphics[width=7cm]{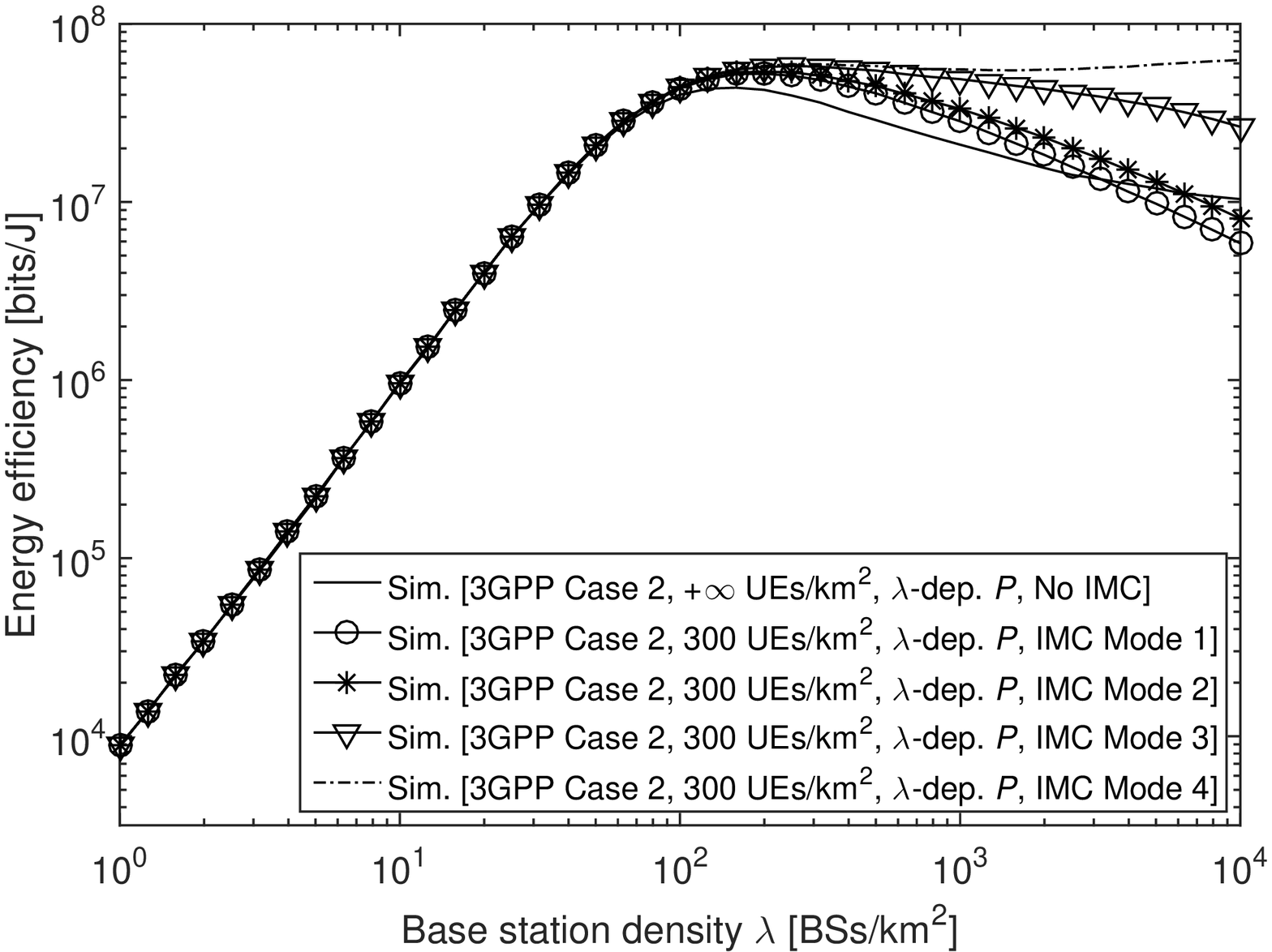}
\label{fig:EE_vs_lambda_gamma0dB_UEdensity300_varP_3GPP_case2}}
\caption{The EE performance with the BS density dependent power configuration and various IMC modes.}
\label{fig:EE_performance}
\vspace{-0.3cm}
\end{figure*}

\subsection{The EE Performance\label{subsec:the_EE_performance}}

As discussed in Subsection~\ref{subsec:The-Energy-Efficiency}, since
we consider the realistic EE performance, we should acknowledge the
fact that modern telecommunication systems usually work in the interference
limited regime and the BS transmission power $P$ should vary with
$\lambda$. In this section, we formulate $P$ using the practical
power model presented in~\cite{Tutor_smallcell}. Specifically, the
transmit power of each BS is configured such that it provides a signal-to-noise-ratio
(SNR) of $\eta_{0}=15$\ dB at the edge of the average coverage area
for a UE with NLoS transmissions, which corresponds to the worst-case
path loss. In addition, the distance from a cell-edge UE to its serving
BS with an average coverage area is calculated by $r_{0}=\sqrt{\frac{1}{\lambda\pi}}$,
which is the radius of an equivalent disk-shaped coverage area with
an area size of $\frac{1}{\lambda}$.%
{} Therefore, the worst-case pathloss is given by $A^{{\rm {NL}}}r_{0}^{-\alpha^{{\rm {NL}}}}$
and the required transmission power to enable a $\eta_{0}$\ dB SNR
for this case can be computed as~\cite{Tutor_smallcell}%
\begin{eqnarray}
P\left(\lambda\right) & = & \frac{10^{\frac{\eta_{0}}{10}}P_{{\rm {N}}}}{A^{{\rm {NL}}}r_{0}^{-\alpha^{{\rm {NL}}}}}.\label{eq:power_setting_edge_SNR}
\end{eqnarray}

In Fig.~\ref{fig:BS_den_depend_TxP}, we plot the BS density dependent
transmission power in dBm to illustrate this realistic power configuration
when $\eta_{0}=15$\ dB. Note that our modeling of $P$ is very practical,
covering the cases of macrocells and picocells recommended in the
3GPP Long-Term Evolution (LTE) networks. More specifically, the typical
BS densities of LTE macrocells and picocells are respectively several\ $\textrm{BSs/km}^{2}$
and around 50\ $\textrm{BSs/km}^{2}$~\cite{TR36.814}. As a result,
the typical $P$ of macrocell BSs and picocells BSs are respectively
assumed to be 46\ dBm and 24\ dBm in the 3GPP standards~\cite{TR36.814},
which match well with our modeling of $P$ in Fig.~\ref{fig:BS_den_depend_TxP}.%

As a result of (\ref{eq:power_setting_edge_SNR}), $P_{{\rm {IMC}}}^{{\rm {TOT}}}\left(\lambda\right)$
and $P_{{\rm {ACT}}}^{{\rm {TOT}}}\left(\lambda\right)$ in (\ref{eq:EE_def})
are calculated numerically using the Green-Touch power model~\cite{DessetPowerModelling},
and the results are displayed in Fig.~\ref{fig:BS_den_depend_totP}
assuming a future SCN BS model in year 2020 and a 10$\,$MHz bandwidth.
From this figures, we can draw the following observations:
\begin{itemize}
\item The total power of each active BS, i.e., $P_{{\rm {ACT}}}^{{\rm {TOT}}}\left(\lambda\right)$,
is always larger than that of each idle BS, i.e., $P_{{\rm {IMC}}}^{{\rm {TOT}}}\left(\lambda\right)$,
because some BS component(s) will be deactivated to save energy consumption
when a BS enters an idle mode.
\item As mentioned in Subsection~\ref{subsec:The-Energy-Efficiency}, we
consider \emph{the Green-Touch slow idle mode} and \emph{the Green-Touch
shut-down mode} to characterize $P_{{\rm {IMC}}}^{{\rm {TOT}}}\left(\lambda\right)$,
which are represented by IMC Mode~1 and IMC Mode~2, respectively.
In comparison, IMC Mode~2 consumes less energy than IMC Mode~1 as
shown in Fig.~\ref{fig:BS_den_depend_totP}.
\item Following~\cite{Tutor_smallcell}, we also consider two futuristic
idle modes to further characterize $P_{{\rm {IMC}}}^{{\rm {TOT}}}\left(\lambda\right)$,
where their energy consumption is 15\% (IMC Mode~3) or 1\% (IMC Mode~4)
of that of \emph{the Green-Touch slow idle mode} (IMC Mode~1). The
former mode (IMC Mode~3) accounts less energy consumption than \emph{the
Green-Touch shut-down mode} (IMC Mode~2), and the latter mode (IMC
Mode~4) assumes that a BS consumes almost nothing.
\end{itemize}

Based on the results of $P_{{\rm {IMC}}}^{{\rm {TOT}}}\left(\lambda\right)$
and $P_{{\rm {ACT}}}^{{\rm {TOT}}}\left(\lambda\right)$ displayed
in Fig.~\ref{fig:BS_den_depend_totP}, in Fig.~\ref{fig:EE_performance}
we plot the EE performance for 3GPP Case~1 and Case~2 when $\rho=+\infty\,\textrm{UEs/km}^{2}$
without the IMC and $\rho=300\,\textrm{UEs/km}^{2}$ with various
IMC modes.%

Here, Fig.~\ref{fig:EE_vs_lambda_gamma0dB_UEdensity300_varP_3GPP_case1}
shows our analytical results for 3GPP Case~1 based on the ASE performance
exhibited in Subsection~\ref{subsec:Sim-ASE-3GPP-Case-1}, while
Fig.~\ref{fig:EE_vs_lambda_gamma0dB_UEdensity300_varP_3GPP_case2}
displays our simulation results for 3GPP Case~2 based on the ASE
performance discussed in Subsection~\ref{subsec:perfm_3GPP_Case2}.
Although 3GPP Case~2 is more realistic than 3GPP Case~1, as one
can observe from Fig.~\ref{fig:EE_vs_lambda_gamma0dB_UEdensity300_varP_3GPP_case1}
and Fig.~\ref{fig:EE_vs_lambda_gamma0dB_UEdensity300_varP_3GPP_case2},
the EE performance shows the same trend in both figures%
, only with some quantitative deviations. Again, this indicates the
usefulness of our theoretical analysis to predict the network performance
trend for dense SCNs with the BS IMC.

As discussed in Subsection~\ref{subsec:The-Energy-Efficiency}, $\tilde{\lambda}$
represents the active BS density, which is a function of $\lambda$
due to the BS IMC. Hence, $\lambda$ is used as the x-axis instead
of $\tilde{\lambda}$ in Fig.~\ref{fig:EE_performance}.

From Fig.~\ref{fig:EE_performance}, we can draw the following conclusions:
\begin{itemize}
\item As predicted in Subsection~\ref{subsec:Sim-ASE-3GPP-Case-1}, the
baseline scheme with $\rho=+\infty\,\textrm{UEs/km}^{2}$, where all
BSs are active, is the least energy efficient scheme for most BS densities,
because each BS suffers from a diminishing EE return with the network
densification due to the deteriorating performance of the coverage
probability as the BS density increases (see Fig.~\ref{fig:p_cov_vs_lambda_gamma0dB_UEdensity300}).
Such deteriorating performance is caused by the interference path
transition from NLoS to LoS as discussed in previous sections.
\item On the other hand, the EE performance of various IMC modes benefits
from \emph{the Coverage Probability Takeoff}, which improves the performance
of each active BS as the SCN densifies, and thus the IMC scheme outperforms
the baseline scheme in terms of the EE. When comparing the EE performance
of different IMC modes, it can be seen that the lower the power consumption
in the idle mode exhibited in Fig.~\ref{fig:BS_den_depend_totP},
the larger the EE of such IMC mode.%
{}
\item When using \emph{the Green-Touch slow idle mode} (IMC Mode~1) and
\emph{the Green-Touch shut-down mode} (IMC Mode~2), the EE first
increases and then decreases with the network densification%
. This decrease is because the increase in the ASE provided by \emph{the
Coverage Probability Takeoff} is not large enough to compensate the
increase in power consumption that the dense network brings about,
mostly because idle BSs following the Green-Touch power models still
consume a non-negligible amount of energy. Nevertheless, the schemes
with the IMC are superior to the baseline scheme. In more detail,
when $\lambda=10^{3}\,\textrm{BSs/km}^{2}$, \emph{the Green-Touch
slow idle mode} (IMC Mode~1) and \emph{the Green-Touch shut-down
mode} (IMC Mode~2) can achieve EE performance of $17.2\,\textrm{Mbits/J}$
and $20.2\,\textrm{Mbits/J}$, respectively, which are around two
times the EE of the baseline scheme, i.e., $9.95\,\textrm{Mbits/J}$.
\item When considering the EE of the futuristic IMC Mode~3 and IMC Mode~4,
the above trend starts changing. For IMC Mode~3, the EE is always
larger than that of the baseline scheme across all BS densities, as
BSs consume much less energy in this idle mode. For IMC Mode~4, idle
BSs barely consume any energy, and thus the above trend fundamentally
alters, i.e., %
as the network evolves into an ultra-dense one, the EE continuously
increases. As a result, when $\lambda=10^{3}\,\textrm{BSs/km}^{2}$,
IMC Mode~3 and IMC Mode~4 can achieve EE performance of $29.6\,\textrm{Mbits/J}$
and $33.6\,\textrm{Mbits/J}$, respectively, which triple that of
the baseline scheme, i.e., $9.95\,\textrm{Mbits/J}$. This help us
to conclude that idle mode schemes similar to IMC Mode~4 are needed
to ensure an energy-efficient deployment of dense SCNs in 5G and beyond.
\end{itemize}

\subsection{Future Work of Ultra-Dense SCNs\label{subsec:future_work}}

In this subsection, we indicate several research directions for ultra-dense
SCNs:
\begin{itemize}
\item It would be good to study a proportional fair (PF) scheduler in ultra-dense
networks~\cite{Jafari2015scheduling}. Currently, in stochastic geometry
analyses, usually a typical UE is randomly chosen for the performance
analysis, which implies that a round Robin (RR) scheduler is employed
in each BS. However, in the 3GPP performance evaluations, the typical
UE is not chosen randomly and a PF scheduler is often used as an appealing
scheduling technique to smartly serve UEs that can offer a better
system throughput than the RR scheduler.
\item It would be good to study the near-field effect in the context of
ultra-dense networks. In particular, the Rayleigh distance as investigated
in~\cite{Wu2014NF}, should be considered in the extremely ultra-dense
networks because the BS-to-UE distance becomes very small as the network
densifies.
\item A very recent discovery shows the 5G network capacity might decrease
to \emph{zero} if the antenna height difference between BSs and UEs
is \emph{non-zero}~\cite{Ding2016GC_ASECrash}. Hence, it is of great
interest to study whether the BS IMC can help to mitigate such network
capacity crash.
\item It would be good to study a non-uniform distribution of BSs with some
constraints on the minimum BS-to-BS distance~\cite{Our_DNA_work_TWC15}.
In stochastic geometry analyses, BSs are usually assumed to be uniformly
deployed in the interested network area. However, in the 3GPP performance
evaluations, small cell clusters are often considered, and it is forbidden
to place any two BSs too close to each other. Such assumption is in
line with the realistic network planning to avoid strong inter-cell
interference.
\item It would be good to study ultra-dense networks in new emerging network
scenarios, such as heterogeneous networks~\cite{Zhang2016UDNhetnet},
distributed networks~\cite{Ge2016UDNs}, high mobility applications~\cite{Ge2016UEmob,Arshad2016UDNhandover},
device to device (D2D) communications~\cite{Liu2014D2D,Liu2015D2Dsurvey,Liu2016D2D},
body area networks~\cite{Sun2016BAN}, unmanned aerial vehicles~\cite{pre_work_drone_BSs_GC16},
etc.
\end{itemize}

\section{Conclusion\label{sec:Conclusion}}

In this paper, we have studied the performance impact of the IMC on
dense SCNs considering probabilistic LoS and NLoS transmissions. The
impact is significant on the coverage probability performance, i.e.,
as the BS density surpasses the UE density, the coverage probability
continuously increases toward one in dense SCNs \emph{(the Coverage
Probability Takeoff}), addressing the critical issue of coverage probability
decrease that may lead to ``the death of 5G''.

Two important conclusions have been drawn from our study: (i) the
active BS density with the mentioned probabilistic LoS and NLoS path
loss model is lower-bounded by that with a simplistic single-slope
path loss model derived in~\cite{dynOnOff_Huang2012}, and (ii) such
lower bound, shown in~\cite{dynOnOff_Huang2012}, is tight, especially
for dense SCNs. This shows a simple way of studying the IMC in dense
SCNs.

Moreover, from our studies based on practical power models of the
Green-Touch project and realistic 3GPP propagation models, we conclude
that %
idle mode schemes similar to IMC Mode~4 are needed to ensure an energy-efficient
deployment of dense SCNs in 5G and beyond.%

\section*{Appendix~A: Proof of Lemma~\ref{lem:larger-CP}\label{sec:Appendix-newA}}

To prove Lemma~\ref{lem:larger-CP}, first we would like to emphasize
the insights or the proof sketch of Theorem~\ref{thm:p_cov_UAS1}
as follows. In (\ref{eq:Theorem_1_p_cov}), $T_{n}^{{\rm {L}}}$ and
$T_{n}^{{\rm {NL}}}$ are the components of the coverage probability
for the case when the signal comes from \emph{the $n$-th piece LoS
path} and for the case when the signal comes from \emph{the $n$-th
piece NLoS path}, respectively. The calculation of $T_{n}^{{\rm {L}}}$
is based on (\ref{eq:geom_dis_PDF_UAS1_LoS_thm}) and (\ref{eq:Pr_SINR_req_UAS1_LoS_thm}),
which are explained in the sequel.
\begin{itemize}
\item In (\ref{eq:geom_dis_PDF_UAS1_LoS_thm}), $f_{R,n}^{{\rm {L}}}\left(r\right)$
characterizes the geometrical density function of the typical UE with
\emph{no other LoS BS} and \emph{no NLoS BS} providing a better link
to the typical UE than its serving BS (a BS with \emph{the $n$-th
piece LoS path}).
\item In (\ref{eq:Pr_SINR_req_UAS1_LoS_thm}), $\exp\left(-\frac{\gamma P_{{\rm {N}}}}{P\zeta_{n}^{{\rm {L}}}\left(r\right)}\right)$
is the probability that \emph{the signal power exceeds the noise power}
by a factor of at least $\gamma$, and $\mathscr{L}_{I_{{\rm {agg}}}}^{{\rm {L}}}\left(\frac{\gamma}{P\zeta_{n}^{{\rm {L}}}\left(r\right)}\right)$
(further computed by (\ref{eq:laplace_term_LoS_UAS1_general_seg_thm}))
is the probability that \emph{the signal power exceeds the aggregate
interference power} by a factor of at least $\gamma$.
\item Since $h$ follows an exponential distribution, the product of the
above probabilities yields the probability that \emph{the signal power
exceeds the sum power of the noise and the aggregate interference}
by a factor of at least $\gamma$.
\end{itemize}

The calculation of $T_{n}^{{\rm {NL}}}$ is based on (\ref{eq:geom_dis_PDF_UAS1_NLoS_thm})
and (\ref{eq:Pr_SINR_req_UAS1_NLoS_thm}). The interpretation of (\ref{eq:geom_dis_PDF_UAS1_NLoS_thm})
and (\ref{eq:Pr_SINR_req_UAS1_NLoS_thm}) are similar to that for
the calculation of $T_{n}^{{\rm {L}}}$.

Hence, Lemma~\ref{lem:larger-CP} is valid because:
\begin{itemize}
\item For $p^{\textrm{cov}}\left(\lambda,\gamma\right)$ with the BS IMC
and that with all BSs being active, (\ref{eq:geom_dis_PDF_UAS1_LoS_thm})
and (\ref{eq:geom_dis_PDF_UAS1_NLoS_thm}) are the same, indicating
an increasing signal power as $\lambda$ grows. This is because that
as $\lambda$ increases, to achieve the same $f_{R,n}^{{\rm {L}}}\left(r\right)$
in (\ref{eq:geom_dis_PDF_UAS1_LoS_thm}) or $f_{R,n}^{{\rm {NL}}}\left(r\right)$
in (\ref{eq:geom_dis_PDF_UAS1_NLoS_thm}), $r$ has to be reduced,
meaning that the typical UE will connect to a nearer BS with a larger
signal power.
\item For $p^{\textrm{cov}}\left(\lambda,\gamma\right)$ with the BS IMC,
$\tilde{\lambda}$ is plugged into (\ref{eq:laplace_term_LoS_UAS1_general_seg_thm})
and (\ref{eq:laplace_term_NLoS_UAS1_general_seg_thm}), while for
$p^{\textrm{cov}}\left(\lambda,\gamma\right)$ with all BSs being
active, $\lambda$ was used in (\ref{eq:laplace_term_LoS_UAS1_general_seg_thm})
and (\ref{eq:laplace_term_NLoS_UAS1_general_seg_thm})~\cite{our_work_TWC2016}.
The former case is able to generate a larger $p^{\textrm{cov}}\left(\lambda,\gamma\right)$
than the latter one, since $\tilde{\lambda}\leq\lambda$ and $\exp\left(-x\right)$
is a decreasing function with respect to $x$ in (\ref{eq:laplace_term_LoS_UAS1_general_seg_thm})
and (\ref{eq:laplace_term_NLoS_UAS1_general_seg_thm}). The intuition
is that the aggregate interference power of the former case with the
BS IMC is less than that of the latter case without, since $\mathscr{L}_{I_{{\rm {agg}}}}^{{\rm {L}}}\left(\frac{\gamma}{P\zeta_{n}^{{\rm {L}}}\left(r\right)}\right)$
in (\ref{eq:laplace_term_LoS_UAS1_general_seg_thm}) and $\mathscr{L}_{I_{{\rm {agg}}}}^{{\rm {NL}}}\left(\frac{\gamma}{P\zeta_{n}^{{\rm {NL}}}\left(r\right)}\right)$
in (\ref{eq:laplace_term_NLoS_UAS1_general_seg_thm}) capture the
impact of the aggregate interference on $p^{\textrm{cov}}\left(\lambda,\gamma\right)$,
as discussed above.
\end{itemize}

\section*{Appendix~B: Proof of Theorem~\ref{thm:lambda_tilde_LB}\label{sec:Appendix-newB}}

For clarity, the main idea of our proof is summarized as follows:
\begin{itemize}
\item We will prove that from a typical UE's point of view, the equivalent
BS density of the considered UAS based on probabilistic LoS and NLoS
transmissions is larger than that of the nearest-distance UAS based
on single-slope path loss transmissions.
\item Considering such increased equivalent BS density and the fact that
a larger $\lambda$ always leads to a larger $\tilde{\lambda}$ due
to a higher BS diversity, we can conclude that $\tilde{\lambda}\geq\tilde{\lambda}^{{\rm {minDis}}}$.
\end{itemize}

First, let us consider a baseline scenario that all BSs only have
NLoS links to UEs. In such scenario, the nearest-distance UAS is a
reasonable one and the active BS density should be characterized by
$\tilde{\lambda}^{{\rm {minDis}}}$~\cite{dynOnOff_Huang2012}.

Next, for the proposed scenario with probabilistic LoS and NLoS transmissions,
we consider a typical UE $k$ and an arbitrary BS $b$ located at
a distance $r$ from UE $k$. Due to probabilistic LoS and NLoS transmissions,
such BS $b$ can be virtually split into two probabilistic BSs, i.e.,
a LoS BS $b^{{\rm {L}}}$ to UE $k$ with a probability of $\textrm{Pr}^{\textrm{L}}\left(r\right)$
and a NLoS BS $b^{{\rm {NL}}}$ to UE $k$ with a probability of $\left(1-\textrm{Pr}^{\textrm{L}}\left(r\right)\right)$.
Compared with the baseline scenario that all BSs only have NLoS links
to UEs, the equivalent distance from the NLoS BS $b^{{\rm {NL}}}$
to UE $k$ remains to be $r$, while that from the LoS BS $b^{{\rm {L}}}$
to UE $k$ can be calculated as $r_{1}=\underset{r_{1}}{\arg}\left\{ \zeta^{{\rm {NL}}}\left(r_{1}\right)=\zeta^{{\rm {L}}}\left(r\right)\right\} $,
which is shown in (\ref{eq:def_r_1}). The calculation of $r_{1}$
is straightforward because it finds an equivalent position for the
LoS BS $b^{{\rm {L}}}$ as if the LoS transmission is replaced with
a NLoS one. Since a LoS transmission is always stronger than a NLoS
one, we have $r_{1}<r$.

Consequently, in a disk area centered on UE $k$ with a radius of
$r_{1}$, the equivalent BS number is increased by at least $\textrm{Pr}^{\textrm{L}}\left(r\right)$,
which is a non-negative value. Due to the arbitrary value of $r_{1}$,
from a typical UE's point of view, the equivalent BS density of the
considered UAS based on probabilistic LoS and NLoS transmissions is
larger than that of the nearest-distance UAS based on single-slope
path loss transmissions. In other words, the existence of LoS BSs
provides more candidate BSs for a typical UE to connect with, and
thus the equivalent BS density increases for each UE.

Finally, we can conclude that $\tilde{\lambda}\geq\tilde{\lambda}^{{\rm {minDis}}}\approx\lambda_{0}\left(q\right)$,
because a larger $\lambda$ leads to a larger $\tilde{\lambda}$ due
to a higher BS diversity.

\section*{Appendix~C: Proof of Theorem~\ref{thm:lambda_tilde_UB}\label{sec:Appendix-newC}}

For clarity, the main idea of our proof is summarized as follows:
\begin{itemize}
\item First, we derive an conditional probability that an arbitrary UE $w$
is \emph{not} associated with an arbitrary BS $b$ conditioned on
the distance between UE $w$ and BS $b$ being $r$. Such conditional
probability is denoted by ${\rm {Pr}}\left[\left.w\nsim b\right|r\right]$.
\item Next, we derive an unconditional probability that an arbitrary UE
$w$ is \emph{not} associated with an arbitrary BS $b$ by performing
an integral over $r$ considering the uniform distribution of UEs
in the considered network. Such unconditional probability is denoted
by ${\rm {Pr}}\left[w\nsim b\right]$.
\item Finally, we derive a lower bound of the probability that every UE
is \emph{not} associated with an arbitrary BS $b$, so that BS $b$
should switch off its transmission. The lower bound of the BS deactivation
probability is then translated to an upper bound of the active BS
density, i.e., $\tilde{\lambda}$.
\end{itemize}

For convenience, the PDF of the distance between a typical UE and
its serving BS, i.e., $\left\{ f_{R,n}^{\textrm{L}}\left(r\right)\right\} $
and $\left\{ f_{R,n}^{\textrm{NL}}\left(r\right)\right\} $ are stacked
into piece-wise functions written as
\begin{equation}
f_{R}^{Path}\left(r\right)=\begin{cases}
f_{R,1}^{Path}\left(r\right), & \textrm{when }0\leq r\leq d_{1}\\
f_{R,2}^{Path}\left(r\right),\hspace{-0.3cm} & \textrm{when }d_{1}<r\leq d_{2}\\
\vdots & \vdots\\
f_{R,N}^{Path}\left(r\right), & \textrm{when }r>d_{N-1}
\end{cases},\label{eq:general_fr_func}
\end{equation}
where the string variable $Path$ takes the value of ``L'' and ``NL''
for the LoS and the NLoS cases, respectively.

Based on $f_{R}^{Path}\left(r\right)$, we define the cumulative distribution
function (CDF) of $r$ as
\begin{equation}
F_{R}^{Path}\left(r\right)=\int_{0}^{r}f_{R}^{Path}\left(v\right)dv.\label{eq:general_Fr_func}
\end{equation}
In addition, we define the sum of $F_{R}^{{\rm {L}}}\left(r\right)$
and $F_{R}^{{\rm {NL}}}\left(r\right)$ as $F_{R}\left(r\right)=F_{R}^{{\rm {L}}}\left(r\right)+F_{R}^{{\rm {NL}}}\left(r\right)$,
which is the CDF of the UE association distance of the presented UAS.
Obviously, we have $F_{R}\left(+\infty\right)=1$. Then, ${\rm {Pr}}\left[\left.w\nsim b\right|r\right]$
can be calculated by (\ref{eq:Pr_w_notAss_b_cond_r_thm}) because
${\rm {Pr}}\left[\left.w\nsim b\right|r\right]$ should be the sum
of the probabilities of the following two events that lead to the
event $\left[\left.w\nsim b\right|r\right]$:
\begin{itemize}
\item The first term of (\ref{eq:Pr_w_notAss_b_cond_r_thm}): The link between
UE $w$ and BS $b$ is a LoS one with a probability of ${\rm {Pr}}^{{\rm {L}}}\left(r\right)$
while UE $w$ is associated with another LoS/NLoS BS that is stronger
than BS $b$ with a probability of $\left[F_{R}^{{\rm {L}}}\left(r\right)+F_{R}^{{\rm {NL}}}\left(r_{1}\right)\right]$,
with $F_{R}^{{\rm {L}}}\left(r\right)$ and $F_{R}^{{\rm {NL}}}\left(r_{1}\right)$
corresponding to the cases of a stronger LoS BS and a stronger NLoS
BS, respectively;
\item The second term of (\ref{eq:Pr_w_notAss_b_cond_r_thm}): The link
between UE $w$ and BS $b$ is a NLoS one with a probability of $\left[1-{\rm {Pr}}^{{\rm {L}}}\left(r\right)\right]$
while UE $w$ is associated with another LoS/NLoS BS that is stronger
than BS $b$ with a probability of $\left[F_{R}^{{\rm {L}}}\left(r_{2}\right)+F_{R}^{{\rm {NL}}}\left(r\right)\right]$,
with $F_{R}^{{\rm {L}}}\left(r_{2}\right)$ and $F_{R}^{{\rm {NL}}}\left(r\right)$
corresponding to the cases of a stronger LoS BS and a stronger NLoS
BS, respectively.
\end{itemize}

Next, for an arbitrary BS $b$, we suppose that all its candidate
UEs are randomly distributed in a disk $\Omega$ centered on BS $b$
with a radius of $r_{\textrm{max}}>0$. Then, for an arbitrary UE
$w$ inside the disk $\Omega$, ${\rm {Pr}}\left[w\nsim b\right]$
can be computed by (\ref{eq:Pr_w_notAss_b_thm}), where $\frac{2r}{r_{\textrm{max}}^{2}}$
is the distribution density function with respect to $r$ for UE $w$~\cite{Jeff2011},
because UEs are assumed to be uniformly distributed.%

Finally, the number of candidate UEs inside disk $\Omega$, denoted
by $K$, should follow a Poisson distribution with a parameter of
$\lambda_{\Omega}=\rho\pi r_{\textrm{max}}^{2}$. Thus, the probability
mass function (PMF) of $K$ can be written as~\cite{Book_Proakis}
\begin{equation}
f_{K}\left(k\right)=\frac{\lambda_{\Omega}^{k}e^{-\lambda_{\Omega}}}{k!},\quad k\in\left\{ 0,1,2,\dots,\right\} .\label{eq:PMF_UE_num_in_Omega}
\end{equation}
Hence, the probability that BS $b$ should be muted, i.e., no UE is
associated with BS $b$, can be computed by (\ref{eq:Q_off_thm}).%

It is very important to note that (\ref{eq:Q_off_thm}) ignores the
correlation between nearby UEs inside disk $\Omega$, i.e., a UE \emph{$k$
not} associated with BS $b$ may imply that a nearby UE $k'$ should
have a large probability of also \emph{not} connecting with BS $b$,
due to the possible existence of a high-link-quality BS near UEs \emph{$k$
}and\emph{ }$k'$. Therefore, $Q^{{\rm {off}}}$ under-estimates the
probability that BS $b$ should be muted, and thus the active BS density
$\tilde{\lambda}$ can be upper-bounded by $\lambda\left(1-Q^{{\rm {off}}}\right)$,
which concludes our proof.

\bibliographystyle{IEEEtran}
\bibliography{Ming_library}

\end{document}